\documentclass[12pt, a4paper, oneside]{book}

\usepackage{amsmath, hyperref, setspace, graphicx, caption, subcaption, fancyhdr}
\usepackage[a4paper, top=1in, bottom=1in, left=1in, right=1in]{geometry}
\graphicspath{ {./02_Images/}}
\usepackage[export]{adjustbox}
\usepackage[style=apa, sorting=nyvt]{biblatex}
\hypersetup{
  colorlinks = true,
  citecolor = black,
  urlcolor = blue,
  linkcolor=black,
  pdfauthor = {Kenil Ajudiya},
  pdftitle = {Searching For Fast Radio Transients And Radio Pulsars Using SPOTLIGHT},
  pdfpagemode = FullScreen,
  bookmarksopen = true}

\newcommand{\chref}[1]{Chapter \ref{#1}}
\newcommand{\secref}[1]{Section \ref{#1}}
\newcommand{\figref}[1]{Figure \ref{#1}}
\newcommand{\rnum}[1]{\uppercase\expandafter{\romannumeral #1\relax}}

\begin{document}
    
    \singlespacing
    \begin{titlepage}
    \begin{center}
        
        \Huge{Searching For Fast Radio Transients\\
        And Radio Pulsars\\
        Using SPOTLIGHT}
        
        \vspace{30pt}
        
        \large{\textit{Submitted in partial fulfilment of\\
        the requirements for the award of the degree of}}
        
        \vspace{15pt}
        
        \LARGE{Master of Science (Research)\\
        in Physics}
        
        \vspace{30pt}
        
        \normalsize{by}
        
        \vspace{10pt}
        
        \large{Kenil Ajudiya}\\
        \normalsize{S.R. No. 11-01-00-10-91-20-1-18711\\
        Undergraduate Programme\\
        Indian Institute of Science}
        
        \vspace{10pt}        
        
        \begin{figure}[!h]%
            \centering
            \subfloat{{\includegraphics[height=3cm]{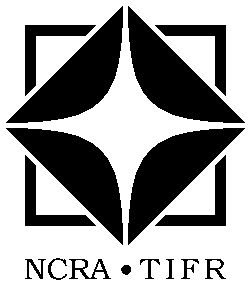}}}%
            \qquad
            \subfloat{{\includegraphics[height=3cm]{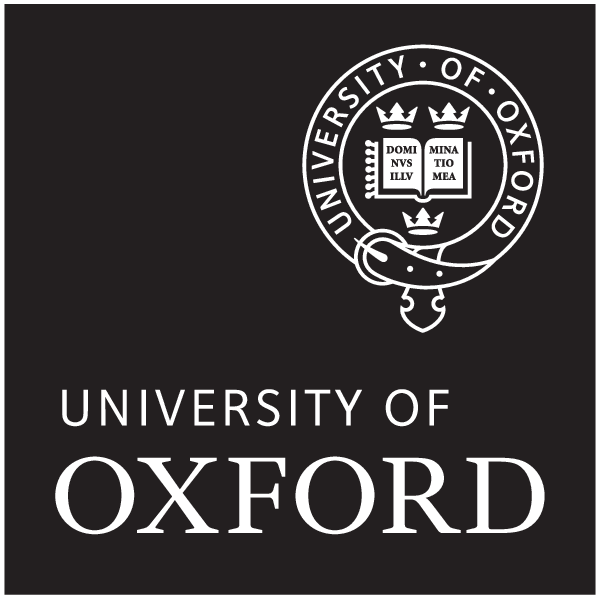}}}%
            \qquad
            \subfloat{{\includegraphics[height=3cm]{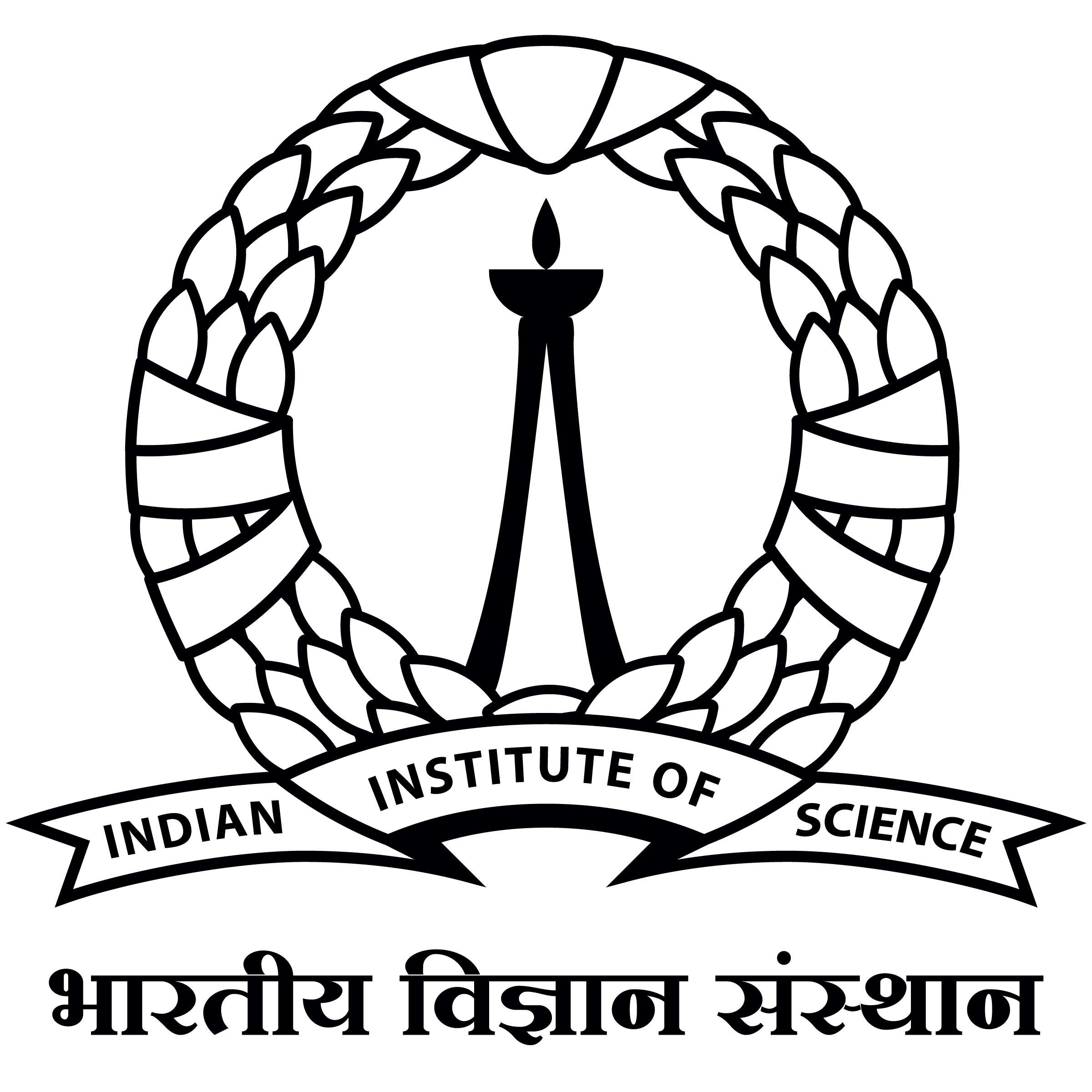}}}%
        \end{figure}

        \vfill

        \normalsize{Under the supervision of}
        
        \vspace{10pt}
        
        \large{Prof. Jayanta Roy}\\
        \normalsize{National Centre for Radio Astrophysics - \\
        Tata Institute of Fundamental Research (NCRA-TIFR), Pune.                                      } 
        
        \vspace{5pt}
        
        \large{Prof. Wesley Armour}\\
        \normalsize{Oxford e-Research Centre (OeRC), Department of Engineering Science, \\
        University of Oxford, Oxford.} 
        
        \vspace{5pt}
        \normalsize{and}
        \vspace{5pt}
        
        \large{Prof. Nirupam Roy}\\
        \normalsize{Department of Physics,\\
        Indian Institute of Science (IISc), Bengaluru.}
        
        \vspace{10pt}
        
        \normalsize{April 24, 2025}
        
    \end{center}
\end{titlepage}
    
    \frontmatter
    \onehalfspacing
    
    \begin{flushright}
    \vspace*{150pt}
    
    Those afraid of the universe as it really is, those who pretend to nonexistent knowledge and envision a Cosmos centered on human beings will prefer the fleeting comforts of superstition. They avoid rather than confront the world. But those with the courage to explore the weave and structure of the Cosmos, even where it differs profoundly from their wishes and prejudices, will penetrate its deepest mysteries.\\
    ---\textsc{Carl Sagan}, \textit{Cosmos}
    
    \vspace*{150pt}
    
    \textbf{Dedicated to my parents,\\
    Mr. and Mrs. Ajudiya}
\end{flushright}
\newpage

\fancyfoot[C]{\thepage}

\chapter*{\centering Acknowledgements}
    \addcontentsline{toc}{chapter}{Acknowledgements}
    Life, at IISc, has changed in countless ways, fortunately, for the better. The role of many people in this change makes me feel grateful and blessed. First and foremost, I would like to thank my mentors Prof. Jayanta Roy, Prof. Wesley Armour, Prof. Nirupam Roy, Prof. Yogesh Wadadekar, Prof. Aloke Kumar, Prof. Chetan Singh Thakur, Prof. Visweshwar Ram Marthi, Dr. Prabu Thiagaraj and Prof. Chandni Usha, all of whom I was fortunate to have worked with on fruitful research endeavours. Lectures on particle physics by Prof. Nirmal Raj, those on the interstellar medium by Prof. Biman Nath, those on quantum mechanics by Prof. Baladitya Suri and those on ecology and animal behaviour by Prof. Rohini Balakrishnan are now memories for me to cherish for the lifetime! I am thankful to many doctoral and graduate students at IISc, at NCRA and at OeRC who have broadened the horizons of my knowledge and understanding through numerous discussions and sometimes, through debates as well. I acknowledge the contributions of the members of the SPOTLIGHT collaboration in the work we collaborated on. Individual mentions can be found in the text at appropriate places.  I am fortunate to have friends like Fida Fathima, Ujjwal Panda, Hemansh Shah, Kruti Bhingradiya, Satyapreet Singh Yadav, Ayushi Chhipa, Balkrishna Sharma and many others, who were there to share the joy in my highs and as a supportive pillar for my mental well-being in my lows. I would like to acknowledge the financial support I received from the grant awarded to the project titled ``Building Indo-UK Collaborations Towards the Square Kilometre Array" funded under the ``DAE-STFC Technology and Skills Programme 2023" initiative, and the Department of science and Technology (DST), Government of India (GOI), via the Kishore Vaigyanik Protsahan Yojna (KVPY). My appreciation also extends to the entire administration staff at IISc, NCRA-TIFR and OeRC. Their collective efforts enabled me to conduct my research efficiently. Lastly, I would like to thank two of the most important people in my life, my mother and my father. None of this would have been possible without their constant support and belief in my pursuits. Thank you for having faith in me and for being there for me as I went through the ups and the downs of my life.

    \vspace{15pt}
    \noindent Kenil Ajudiya\\
    April 13, 2025\\
    Surat, India.
    \newpage
    
    \chapter*{\centering Declaration}
    \addcontentsline{toc}{chapter}{Declaration}
    
    \noindent I, Kenil Ajudiya (SR. No. 1-01-00-10-91-20-1-18711), hereby declare that this thesis entitled ``Searching For Fast Radio Transients And Radio Pulsars Using SPOTLIGHT", submitted in partial fulfilment of the requirements for the award of the degree of Master of Science (Research) in Physics, is a presentation of my original research work done under the guidance and supervision of Prof. Jayanta Roy at the National Centre for Radio Astrophysics - Tata Institute of Fundamental Research (NCRA-TIFR), Pune, Prof. Wesley Armour at Oxford e-Research Centre, University of Oxford, UK, and Prof. Nirupam Roy of the Department of Physics at the Indian Institute of Science (IISc), Bengaluru, during academic year 2024-25, and that it has not been submitted elsewhere for any degree or diploma, conforming to the norms and guidelines given in the Ethical Code of Conduct of all the institutes and organisations involved. Wherever contributions of others are involved, every effort is made to indicate this clearly, with due reference to the literature, and acknowledgement of collaborative research and discussions.
    
    \begin{flushright}
        \includegraphics[scale=0.2]{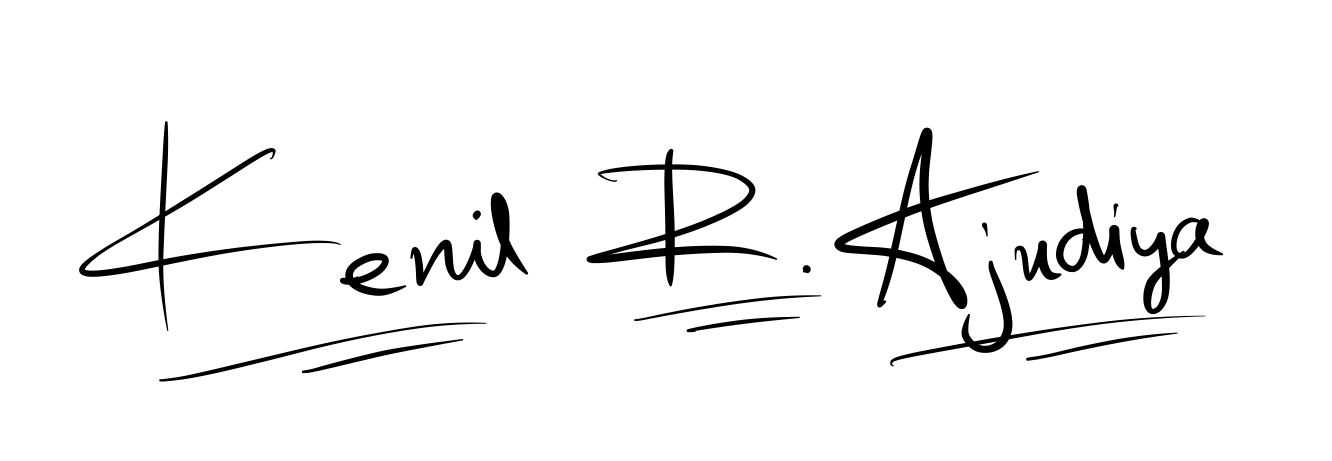}\\
        (Kenil Ajudiya)\hbox{\hskip 25pt}\\
    \end{flushright}
    
    \noindent Place: Surat, India.\\
    \noindent Date: April 13, 2025

\chapter*{\centering Certificate}
    \addcontentsline{toc}{chapter}{Certificate}
    
    \noindent This is to certify that the work contained in this thesis titled ``Searching For Fast Radio Transients And Radio Pulsars Using SPOTLIGHT", submitted by Kenil Ajudiya in partial fulfilment of the requirements for the award of the degree of Master of Science (Research) in Physics, has been carried out by him in part at the National Centre for Radio Astrophysics - Tata Institute of Fundamental Research (NCRA-TIFR), Pune under the supervision of Prof. Jayanta Roy, at Oxford e-Research Centre, University of Oxford, UK under the supervision of Prof. Wesley Armour, and at the Department of Physics at the Indian Institute of Science (IISc), Bengaluru under the supervision of Prof. Nirupam Roy, during academic year 2024-25, and that no part of it has been previously submitted for a degree, diploma or any other qualification at this university or any other institution to the best of my knowledge.
    
    \vspace{1in}
    \begin{flushright}
        (Prof. Jayanta Roy)\hbox{\hskip 25pt}\\
    \end{flushright}
    
    \vspace{1in}
    \begin{flushright}
        (Prof. Nirupam Roy)\hbox{\hskip 25pt}\\
    \end{flushright}
    
    \noindent Date: April 18, 2025
    \newpage
    
    \chapter*{\centering Abstract}
    \addcontentsline{toc}{chapter}{Abstract}
    Our initial impressions of astronomical objects was that they are inherently ``static" over the course of any reasonably long observation. However, with the discovery of quasars and their scintillation in 1963-64, we learnt that there are transient phenomena even at the astronomical scales. The world of known transients has been expanding ever since then. Objects and phenomena like quasars, gamma ray bursts (GRBs), pulsars, rotating radio transients (RRATs), fast radio bursts (FRBs) and ultra long period transients (ULPTs) have answered several unanswered questions about the end states of stellar collapse, i.e, the formation and properties of back holes, neutron stars and white dwarfs. Even more interestingly, they have made us better realise how little we know about the universe. Even after more than 5 decades of research, many lurking questions about neutron stars await answers. In the current work, I explored the arena of FRB and radio pulsar astronomy by joining and contributing to the efforts of the SPOTLIGHT\footnote{It stands for Survey for sPoradic radiO bursTs via a commensaL multI-beam Gpu-powered Hpc at gmrT} collaboration.
    
    The recent decades have witnessed huge leaps in radio instrumentation and high performance computing (HPC) technologies driven by the development of high throughput Graphics Processing Units (GPUs). These major technological advancements are conducive to probing extremely small time scales (up to microseconds) of astronomical events. Modern and next generation radio transients surveys at existing and upcoming radio telescopes worldwide are designed to make optimal use of the available resources to push the research frontiers with the sheer volume of data they produce (hence the terminology, data-driven astronomy). There is an urgent need to upgrade the existing time-domain radio astronomy software to keep up with the pace of the technological revolution on the hardware side.
    
    Although pulsar phenomena has been studied in great detail, several loopholes in our understanding still remain. There are prospects of learning more robust statistical facts about their population to aid theoretical advances. The global race of discovering more FRBs continues to date. In particular, the efforts are stronger towards localising any newly discovered FRBs (both one-off's or repeaters) and discover as many of the repeaters as might be possible. The sheer diversity of these phenomena necessitates a statistically significant population of even the marginalised species to complete the jig-saw puzzle. Moreover, there's always the exciting possibility of finding new puzzles, new phenomena themselves!
    
    SPOTLIGHT is an effort to attack both these issues simultaneously with a powerful weapon, one of the most sensitive radio interferometers at low radio frequencies - the Giant Metrewave Radio Telescope (GMRT). Repeater FRB follow-up capabilities of GMRT have been demonstrated to be cutting-edge. However, being an interferometer restricted to a phased array mode of operation with a maximum of only 4 narrow beams, its survey capabilities were limited in terms of survey speed. SPOTLIGHT overcomes this limitation by tiling the entire primary beam (FoV of a single dish) with $\sim$2000 post-correlation phased array (PCPA)\footnote{It is equivalent to phased array (PA) - incoherent array (IA), thus retaining only the antennae cross-terms in beamforming.} beams. Since the beams are formed from the visibilities with a time-resolution of $\sim$1.3 ms, the visibilities are readily available to image any pulsed emission that might be detected in the beamformed data. Imaging the pulsed signal in real-time thus enables real-time localisation of any detected source. Over the course of 3 years of its operations in commensal search, targetted search and open sky modes, SPOTLIGHT is estimated to detect and localise $\sim$300 FRBs in the frequency range of 300 MHz to 1460 MHz. Following up these discoveries to search for their host galaxies, the contributions of SPOTLIGHT to the number of localised FRBs will increase it many folds from its current value!
    
    This thesis highlights my contributions to the efforts of the software development team of the SPOTLIGHT collaboration. The common aim of the development process is to build software which are capable of handling extremely high data rates and delivering data products ready for cutting-edge scientific scrutiny with minimal human intervention, in pace with the advances in hardware technology.
    \newpage
    
    \tableofcontents
    
    \listoffigures
    
    % \input{01_Chapters/05_abbreviations}
    % \printnomenclature
    
    \mainmatter
    
    \chapter{Introduction}\label{ch1}

    The radio sky has a diverse variety of astrophysical signals emitted from an equally diverse variety of astrophysical sources in an even more mesmerising variety of astrophysical environments. The sources studied in the initial decades of radio astronomy were primarily those which vary significantly only on timescales of millions of years  or longer and hence, are regarded as ``static" throughout the observation duration. The advances in instrumentation and computation made possible the discovery of radio sources that vary on the timescales of milliseconds.
    
\begin{figure}
        \centering
        \includegraphics[width=15cm]{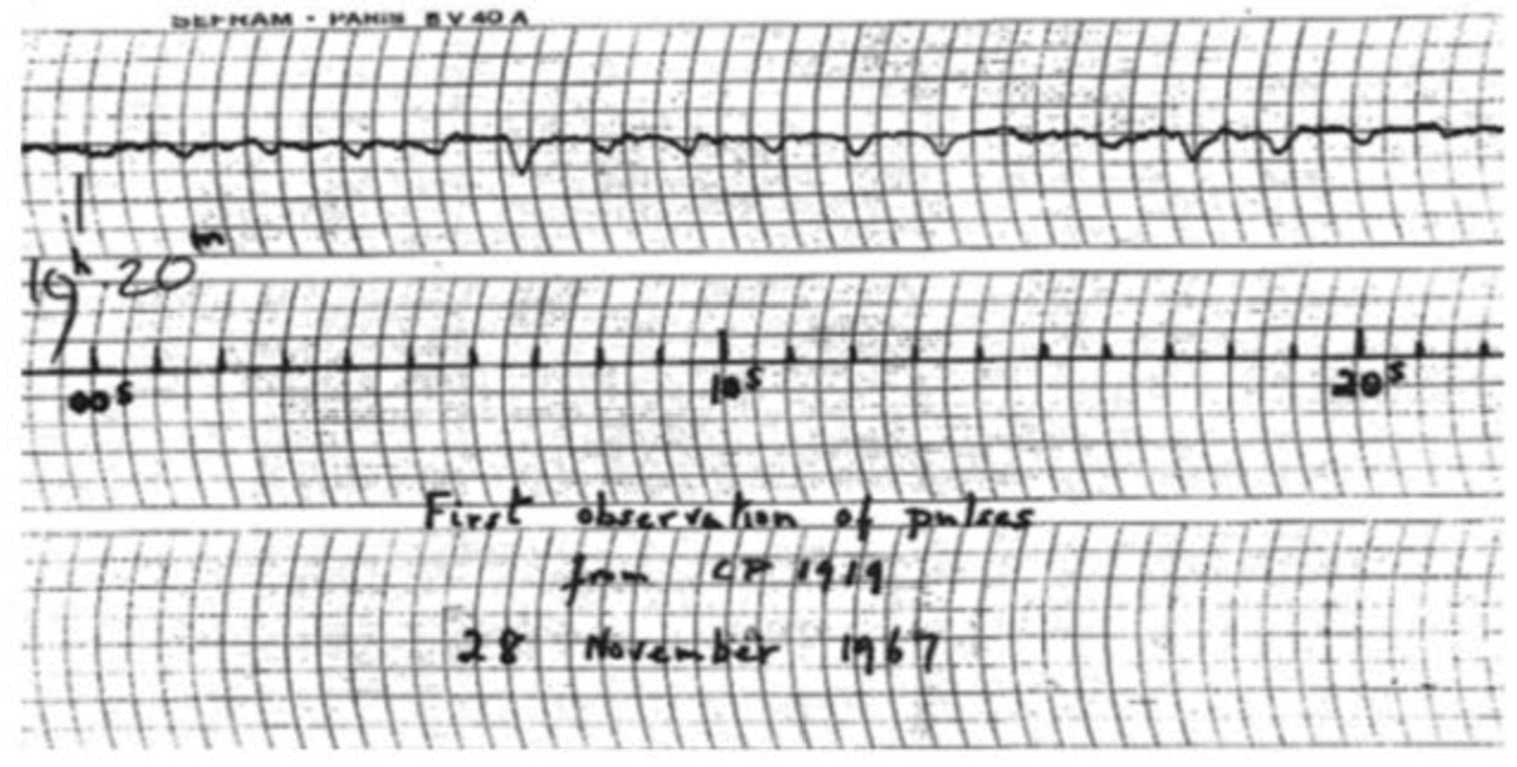}
        \caption{Pen chart recording showing the detection of pulses from the original pulsar B1919+21 first seen during the Cambridge survey (\cite{hewish_observation_1968}). Source: \cite{lorimer_discovery_2024}}
        \label{fig:first_pulsar}
    \end{figure}

    In 1964, the discovery of quasars by \cite{schmidt_3c_1963} and \cite{greenstein_quasi-stellar_1964} and the scintillation of their radio waves as they travel through the ionized interplanetary medium led to events that opened the doors to studying transient phenomena at radio frequencies. The first radio pulsar was then soon discovered, quite serendipitously, in 1967 by \cite{hewish_observation_1968} Jocelyn Bell Burnell, a graduate student working with Prof. Anthony Hewish. While studying quasar scintillation, Bell Burnell saw a signal consisting of regularly spaced pulses, with a separation of exactly 1.337 seconds (see \figref{fig:first_pulsar}), a timescale much shorter than that expected for quasar scintillation. After this discovery, many radio telescopes around the globe started searching for pulsars (henceforth understood to mean radio pulsars, unless stated otherwise), hand-in-hand with further optimizing the instruments and software tools for improved sensitivity at high time-resolution. The following decades saw a boom in the discovery of new pulsars and major theoretical advances in our understanding of their emission mechanism. Pulsar signal carries a signature of the properties of the intervening medium between the observer and the neutron star emitting it. Electromagnetic waves travelling through an ionized medium experiences dispersion (different refractive indices at different frequencies). As a result of dispersion, electromagnetic waves of different frequencies emitted simultaneously arrive at different times at the observer. The difference in the arrival times (or dispersion delay) can be large enough to be measurable at radio frequencies. It is given by
    \begin{equation}
        \begin{gathered}
            \Delta t = t(\nu_1) - t(\nu_2) \approx 4.14881 \times DM \left( \frac{1}{\nu_1^2} - \frac{1}{\nu_2^2}\right)\ \rm{GHz^2cm^3pc^{-1}ms},\\
            DM = \int n_e\ \mathrm{d}l
        \end{gathered}
    \end{equation}
    where $t(\nu)$ is the time of arrival at a given frequency $\nu$, $DM$ is the dispersion measure, $n_e$ is the number density of the free electrons in the intervening medium and the integral is along the line of sight. DM measured from observations of pulsars can be used as an approximate proxy for the column density of electrons between the observer and the source\footnote{See \cite{kulkarni_dispersion_2020} for details and clarification.}. Thus, the DMs of a large enough population of galactic pulsars combined with the measured distances to the neutron stars emitting them can be used to model the galactic free electron density \cite{greiner_tomography_2016}.
    
    Pulsars have a fairly regular pulsed signal, with a well-defined, measurable period and its derivatives. Almost four decades after their discovery, in 2006, a new class of radio sources were discovered by \cite{mclaughlin_transient_2006}, which the authors nicknamed ``rotating radio transients" (RRATs). All the 11 RRATs discovered by \cite{mclaughlin_transient_2006} emitted a very small number of detectable pulses over the course of 35-min search observations, meaning they could not be detected through the widely used Fourier transform search but could be through a search for single pulses. After the discovery of the RRATs, pulsar astronomers world-wide began to use the single-pulse search technique routinely as part of pulsar searches. Despite their sporadicity, the properties of the RRAT single pulses were not significantly different from the properties of normal pulsar single pulses. However, looking at the properties of these objects in this way raised a tantalising question: what other types of objects could exist in this newly revealed single pulse phase space?
    
\begin{figure}%
        \centering
        \subfloat[]{\includegraphics[width=7cm]{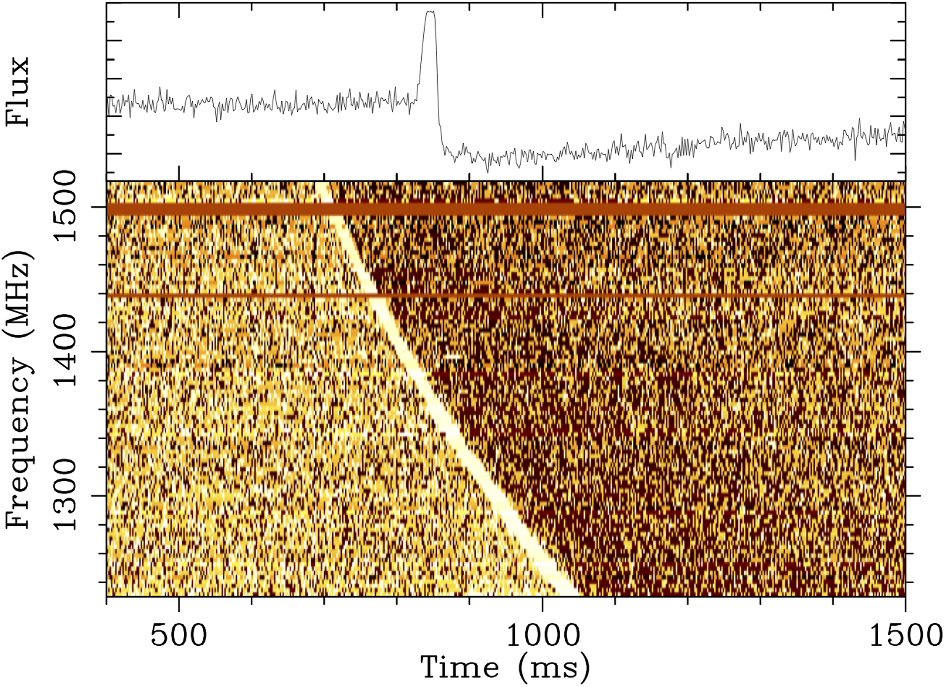}}%
        \qquad
        \subfloat[]{\includegraphics[width=7cm]{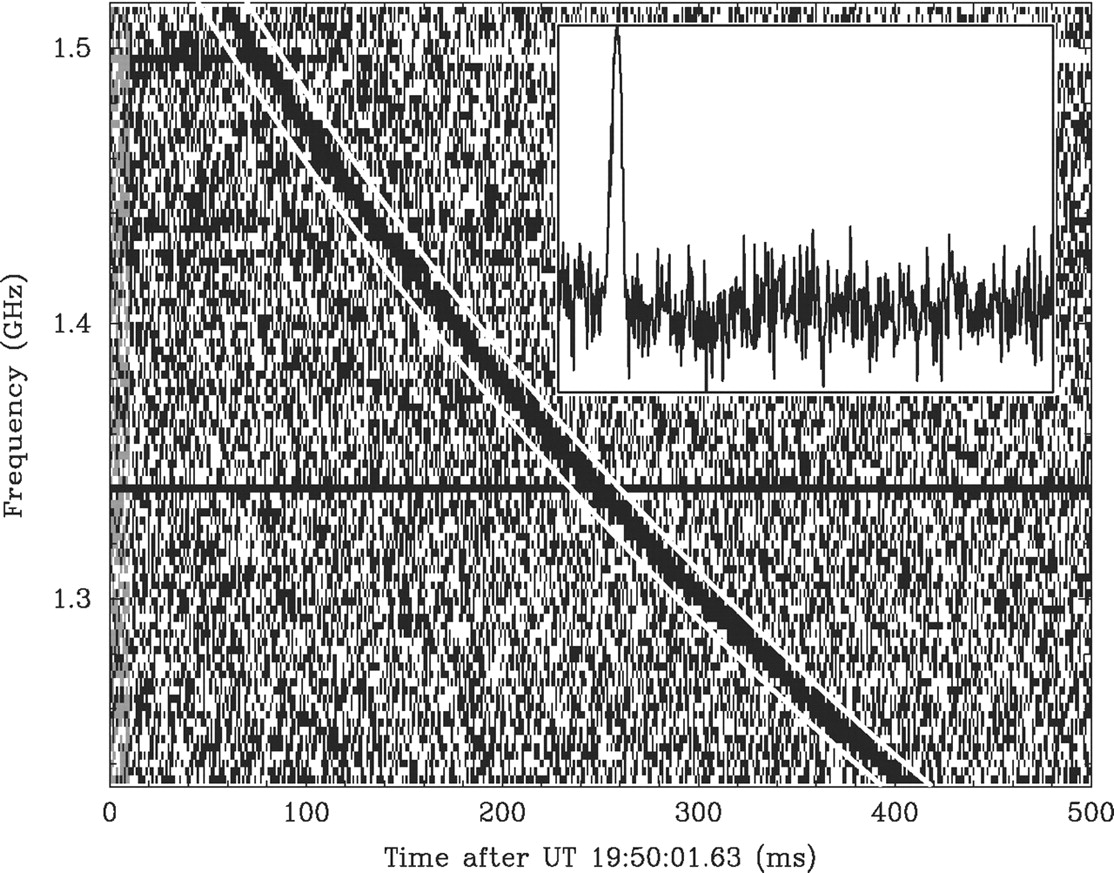}}%
        \caption{(a) The Lorimer burst (\cite{lorimer_bright_2007}), as seen in the beam of the Parkes multi-beam receiver where it appeared brightest, the beam 6. These data have been one-bit digitized and contain 96 frequency channels sampled every millisecond. The burst has a DM of 375 $\rm{cm^{-3}pc}$. The pulse was so bright that it saturated the detector, causing a dip below the nominal baseline of the noise right after the pulse occurred. This signal was also detected in 3 other beams of the receiver. The top panel shows the dedispersed time series and the bottom panel is the dynamic spectrum. The red horizontal lines are frequency channels that have been excised because they are corrupted by RFI. Source: \cite{petroff_fast_2019} (b) The dynamic spectrum of the same burst as seen in the beam 13 where it was initially detected. The inlay shows the dedispersed time series. Source: \cite{lorimer_bright_2007}}%
        \label{fig:Lorimer_burst}%
    \end{figure}

\begin{figure}
        \centering
        \includegraphics[width=15cm]{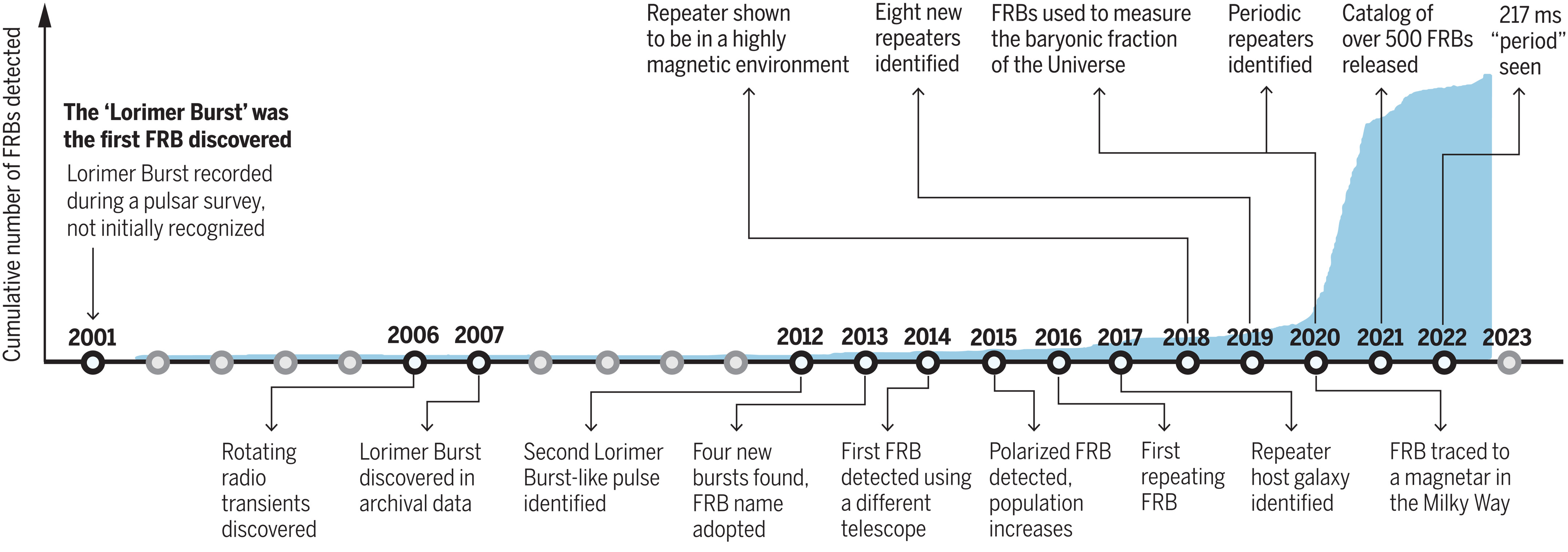}
        \caption{Timeline of some important breakthroughs in FRBs. The blue graph indicates the cumulative number of FRBs detected ($\sim$800, including some bursts from repeaters). Data source: HeRTA: FRBSTATS online catalog. Source: \cite{bailes_discovery_2022}}
        \label{fig:num_FRBs}
    \end{figure}

\begin{figure}
        \centering
        \includegraphics[width=15cm]{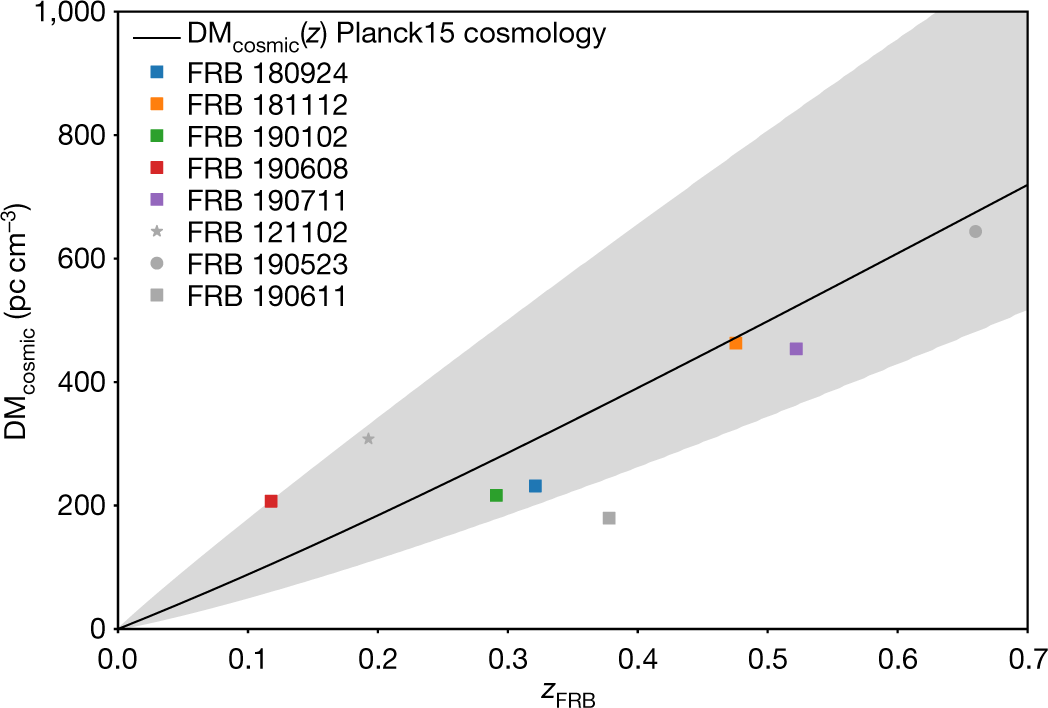}
        \caption{The Macquart relation, as established for the first FRBs with robust host galaxy identifications and redshifts. The correlation between DM and redshift provides a compelling detection of baryons in the intergalactic medium. Credit: J-P Macquart. Source: \cite{lorimer_discovery_2024}}
        \label{fig:Macquart_relation}
    \end{figure}

     The discovery of RRATs was soon followed by another advancement in the field of radio transients - the discovery of a Fast Radio Burst (FRB) (by \cite{lorimer_bright_2007}) in 2007. FRBs are highly luminous enigmatic flashes of radio waves from distant astronomical sources. The first FRB, colloquially known as the ``Lorimer burst", is now also designated FRB 010724 where the numbers represent the date of arrival at earth in the YYMMDD format. It was so bright that it saturated the digitiser levels in the system (see \figref{fig:Lorimer_burst}), was detected in 4 out of the 13 multi-beam pixels of the Parkes Telescope, and the data preparation software had erroneously flagged the signal in beam 6 as interference! After the discovery of the ``Lorimer burst", the global hunt for more FRBs has led to a dramatic increase in the number of known FRBs (see \figref{fig:num_FRBs}); currently many thousands from more than 600 unique sources with almost 50 of them conclusively associated with host galaxies. It is now established beyond any reasonable doubt that the observed FRBs are a part of a cosmological population. Currently, there are many thousands of FRBs known from more than 600 unique sources, with almost 50 of them conclusively associated with host galaxies. With the ongoing boom in FRB discovery and localisation, a significant increase in the sample of FRBs with conclusive host galaxy localisation and thus, measured redshifts is anticipated in the upcoming years. Having gathered a large enough sample, it will be possible to use the DM of the FRBs and their coordinates (RA, Dec and redshift) to model the free electron density of the Universe. The first team to do it was led by J-P Macquart. The result, shown in \figref{fig:Macquart_relation}, is now known as the Macquart relation. The scatter around the Macquart relation contains information about the turbulence of the interstellar gas around galaxy clusters and is now being studied in great detail (\cite{baptista_measuring_2024}). Thus, FRBs can also be used as an independent probe of Cosmology.
    
\begin{figure}%
        \centering
        \subfloat[]{\includegraphics[width=7cm]{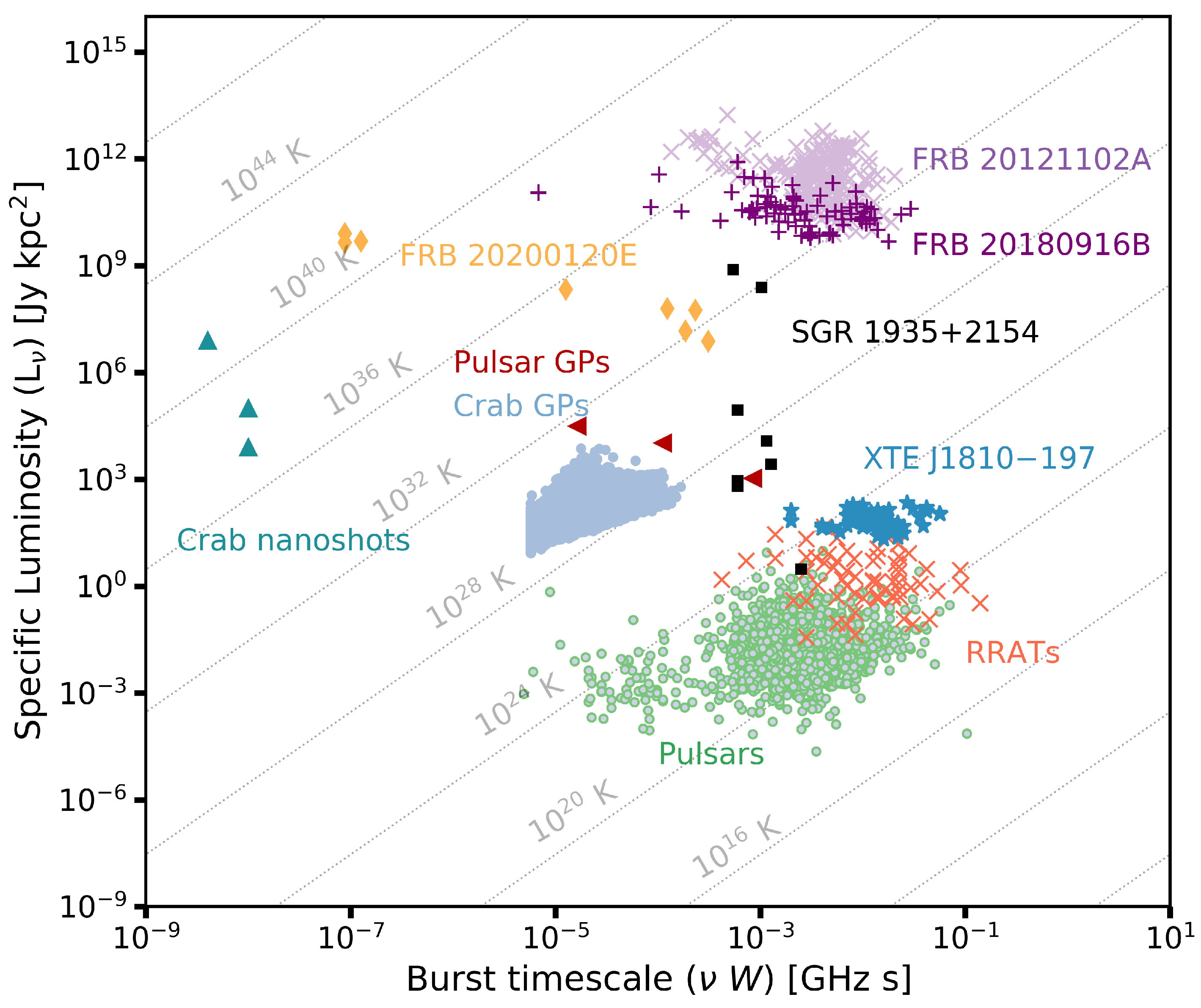}}%
        \qquad
        \subfloat[]{\includegraphics[width=6cm]{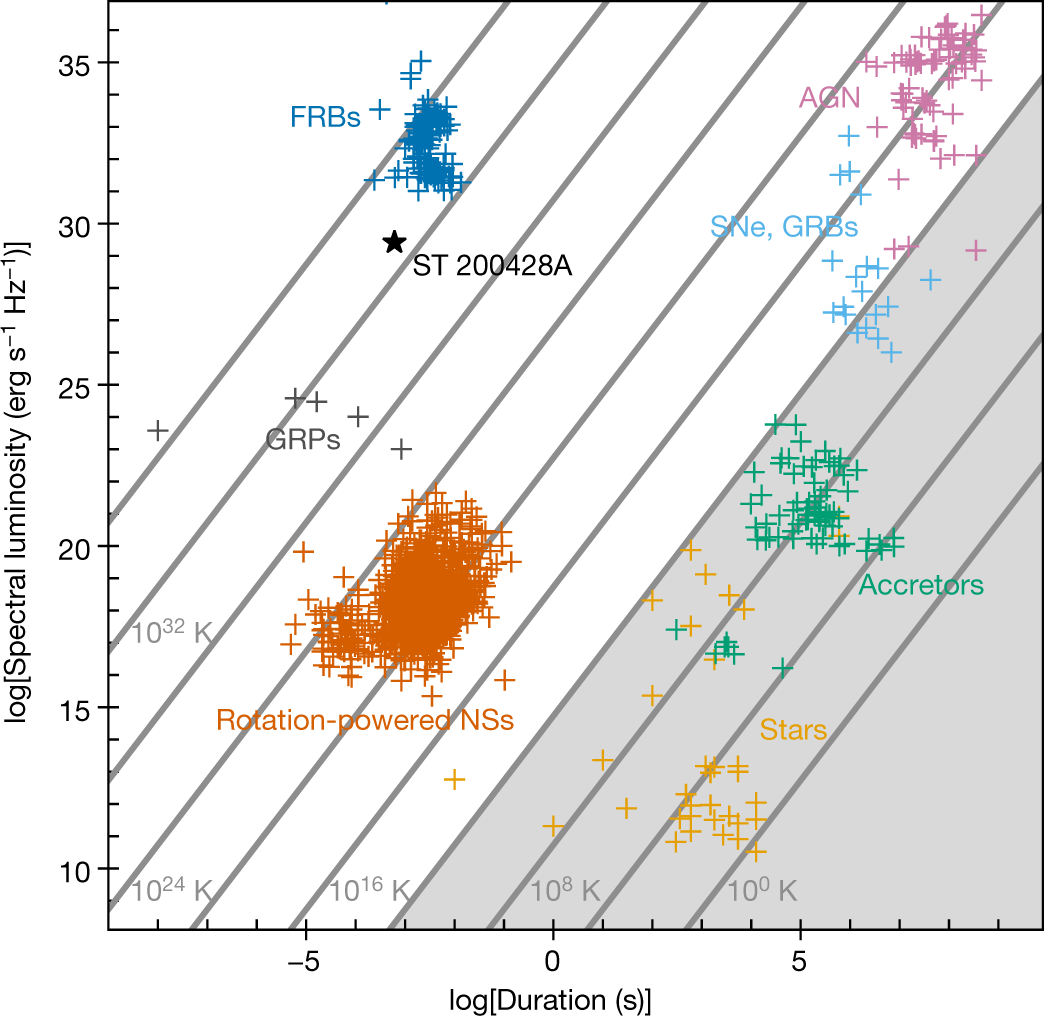}}%
        \caption{The ``luminosity - pulse duration" diagram. Individual pulses from a variety of different sources displayed as luminosity versus pulse width times observing frequency. The Lorimer burst is no longer an anomalously bright object when compared to the other FRBs known. As we note in the text, there is still much parameter space to be explored. Further exciting discoveries in the transient radio sky are anticipated. (a) Source: \cite{caleb_decade_2021} (b) Source: \cite{bochenek_fast_2020}}%
        \label{fig:luminosity-duration}%
    \end{figure}

\begin{figure}%
        \centering
        \subfloat[]{\includegraphics[width=7cm]{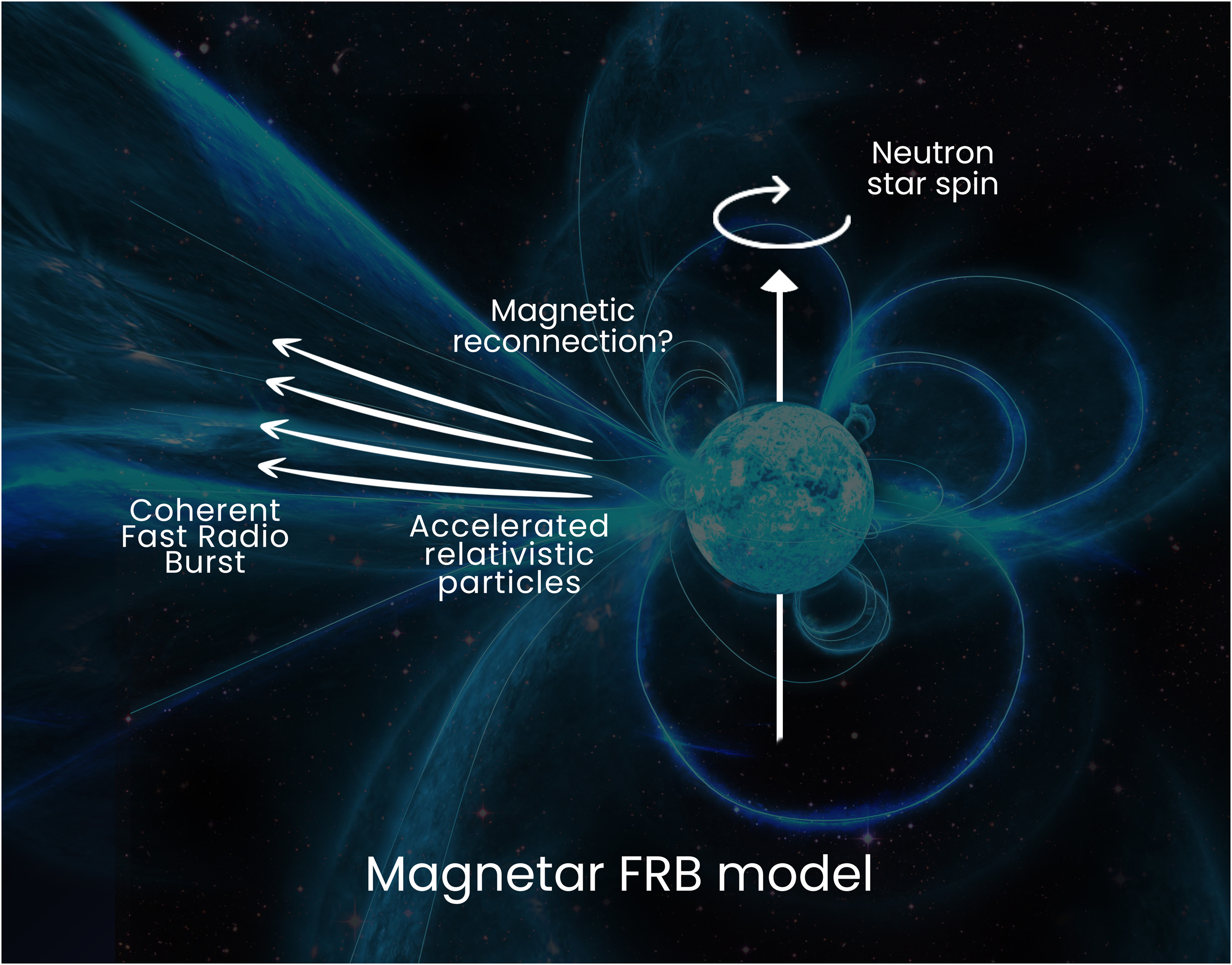}}%
        \qquad
        \subfloat[]{\includegraphics[width=7cm]{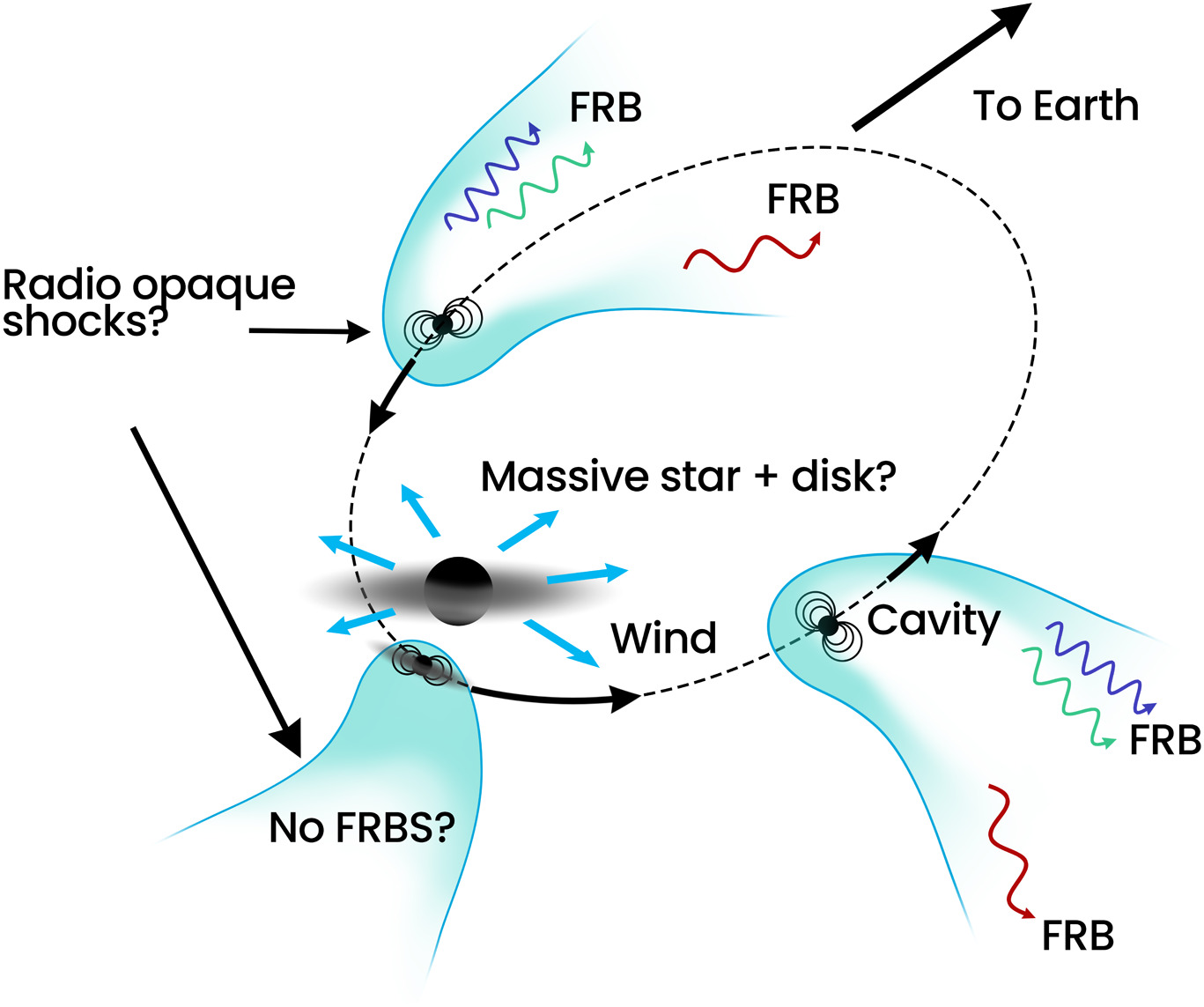}}%
        \caption{(a) Magnetar FRB emission model. Reconfiguration of the intense magnetic fields around a magnetar is associated with high-energy outbursts. In this model for FRB generation, reconfiguration of the magnetic field releases relativistic particles that generate coherent radio emission in the magnetosphere, possibly producing FRBs. Source: \cite{bailes_discovery_2022} (b) Repeating FRB orbital model. In this repeating FRB model (\cite{zhang_cosmic_2017}), a massive star’s stellar wind (blue arrows) causes an orbiting (with a period of weeks to months) magnetar (small black dots) to emit FRBs. The interaction between the stellar wind and the magnetar wind produces cavities (cyan shading) separated by a shock front (cyan lines). The cavity preferentially emits high-frequency FRBs (blue and green wavy arrows) at the leading edge and lower-frequency FRBs (red wavy arrows) in the trailing sections. This model is an attempt to explain both the activity windows of some repeaters and their radio frequency time dependence. Source: \cite{bailes_discovery_2022}}%
        \label{fig:FRB_models}%
    \end{figure}

     In 2016, the FRB 121102 discovered earlier was found to repeat, producing many repeat bursts in a single observation. It was nicknamed the repeater (later known as R1). About 10\% of all currently known sources have been seen to exhibit multiple bursts. It has been shown to be associated with a dwarf galaxy with a high star formation rate at a redshift of $z=0.19$ (\cite{tendulkar_host_2017}), indicating that the burst came from stellar populations such as neutron stars. It was also found that the bursts from the R1 had one of the highest rotation measures ever observed (\cite{michilli_extreme_2018}) which means it came from a very magnetic, dense environment. Among the recent breakthroughs is the discovery of an FRB-like pulse (FRB 200428) from a known magnetar in our Galaxy, SGR 1935+2154. This incredibly bright pulse with a peak flux density of over 1 MJy was spotted by CHIME (CHIME/FRB Collaboration et al. 2020b) and subsequently also by the STARE2 experiment (Bochenek et al. 2020) and quickly localised to a magnetar which was known to be undergoing an enhanced period of activity at the time (April 28, 2020), producing a short X-ray burst contemporaneous with the FRB 200428. Some pulsars, called the giant pulsing pulsars, of which the Crab pulsar is the first-discovered and most well-known example, occasionally emit single pulses that are 100s or even 1000s of times brighter than the average pulse. The estimated luminosity of the pulse was several thousand times brighter than the brightest of the giant pulses, and only a factor of 30 fainter than a typical cosmological FRB pulse (see \figref{fig:luminosity-duration}). It therefore appears to make a strong case for magnetars as plausible sources of FRBs. Magnetars experience magnetic field reconfigurations (see \figref{fig:FRB_models} (a)) which can produce high-energy outbursts by the release of relativistic particles that generate coherent radio emission in the magnetosphere (\cite{coti_zelati_systematic_2018} and \cite{nimmo_magnetospheric_2025}). One of the repeaters, FRB 180916B shows burst activity in a 5-day wide window that recurs every 16.35 days, with a particularly high burst rate in a narrower 0.6-day wide window (\cite{zhang_cosmic_2017}). There are hints of the R1 showing periodic bursting activities with periods of about 150-200 days (\cite{rajwade_possible_2020}, \cite{cruces_repeating_2021}). Periods of this timescale could be indicative of binary motion, and several models have been proposed, binary companion stellar wind models like the one in \figref{fig:FRB_models} (b) being some of the most attractive ones.
    
    The present outlook of the ``luminosity-duration" phase space of radio transient events as shown in \figref{fig:luminosity-duration} exhibits clear grouping of sources with a lot of gaps in the parameter space. Discoveries like FRB 200428 and FRB 20200120E (M81) \footnote{Here, the date of arrival at earth is in the YYYYMMDD format, the letter 'E' refers to the 5th detected event on the mentioned date and ``M81" in the brackets refers to the associated host galaxy.} seems to be a step towards filling some of these gaps. However, we are not currently able to resolve pulses well enough to rule out a class of extremely short duration FRBs (dubbed ultrafast FRBs by some). Although most searched to date are less sensitive to long-duration pulses, Galactic transients with long periods (minutes or more), dubbed ultra-long period transients (ULPTs) by some, are now being discovered (see, for example, \cite{hurley-walker_radio_2022}, \cite{caleb_discovery_2022}, \cite{hurley-walker_long-period_2023}, \cite{caleb_emission-state-switching_2024} and \cite{lee_emission_2025}). The future holds surprises to be discovered; a possible unification scheme for all types of radio transients discussed above (pulsars, RRATs, FRBs, ULPTs and the diversity therein) or perhaps different emission and progenitor models for different types of radio transients.
    
    To better understand these phenomena, there is a need to accelerate the discovery rate, which is possible by the ongoing and upcoming radio transients surveys at radio observatories across the world. These surveys are anticipated to be highly sensitive and generate enormous volumes of data which is unfeasible to be stored. Hence, there is the need to search the data in real-time covering as much of the parameter space as possible. The modern leaps in high-performance computing (HPC) technologies will surely aid the developments. With these, however, arises the need to develop software capable of leveraging the full capabilities of HPC technologies and automating as many tasks as possible. The \href{https://spotlight.ncra.tifr.res.in/}{Survey for sPoradic radiO bursTs via a commensaL multI-beam Gpu-powered Hpc at gmrT (SPOTLIGHT)} is one such attempt, demonstrating the search survey capabilities of the \href{https://www.gmrt.ncra.tifr.res.in/}{Giant Metrewave Radio Telescope (GMRT)}. The SPOTLIGHT project introduces a new mode of radio observations at GMRT - the post-correlation phased array (PCPA) mode - with its multi-beam correlator and beamformer capable of forming up to 2000 PC beams within the primary field of view of a single dish with temporal resolution of 1.31072 ms. The interferometric capabilities of GMRT will readily enable up to arc-minute localisation to one of the 2000 beams for any detected burst since that is the typical angular size of a PCPA beam of GMRT. For all the events positively classified by the CNN-based classifier, FETCH (\cite{agarwal_fetch_2020}), the post correlation visibilities will be used to image the burst to achieve arc-second localisation and estimate its fluence. For all such events, full-resolution, Nyquist-sampled raw (antennae) voltages will also be stored for further, detailed scrutiny. A selected fraction of all the beams will be recorded and searched for pulsars on a weekly basis. With its commensal operations over the course of three years, SPOTLIGHT is anticipating to discover $\sim$300 FRBs and numerous pulsars in the low frequency regime (300 MHz to 1460 MHz), placing GMRT in the league of leading radio transients discovery instruments around the globe. Towards achieving such a feat, the current work aims to contribute to the development of software capabilities for search and localisation of FRBs and pulsars, leveraging the capabilities of the GPU-powered HPC at GMRT.
    
    The work presented in this thesis highlights my contributions to the development and testing of the real-time multi-beam transient search pipeline (a.k.a. FRB detection pipeline) (\chref{ch2}) and the quasi-real-time pulsar search pipeline (\chref{ch3}) to be deployed at GMRT under the SPOTLIGHT project. \chref{ch2} opens with an overview of the multi-beam transient search pipeline and describes my contribution to its interfacing with the new SPOTLIGHT multi-beam beamformer and the efforts to incorporate a real-time multi-beam RFI mitigation software in the interface. It concludes with a description of my contributions to the development of a single pulse injection software and the outcomes of numerous tests of various pipeline components. \chref{ch3} opens with an overview of the pulsar search pipeline and motivates the necessity of a GPU-accelerated folding algorithm. It then presents a GPU-accelerated Fast Folding Algorithm (FFA)-based pulsar folding software that I have been developing during my visit to the Oxford e-Research Centre (OeRC), University of Oxford, UK. The thesis concludes with highlighting the significance of the work and the prospects of future developments in \chref{ch4}.
    
    \chapter{Developing the Transient Search Pipeline}\label{ch2}
    \section{Overview of the transient search pipeline}\label{sec2.1}
\begin{figure}
            \centering
            \includegraphics[width=15.5cm]{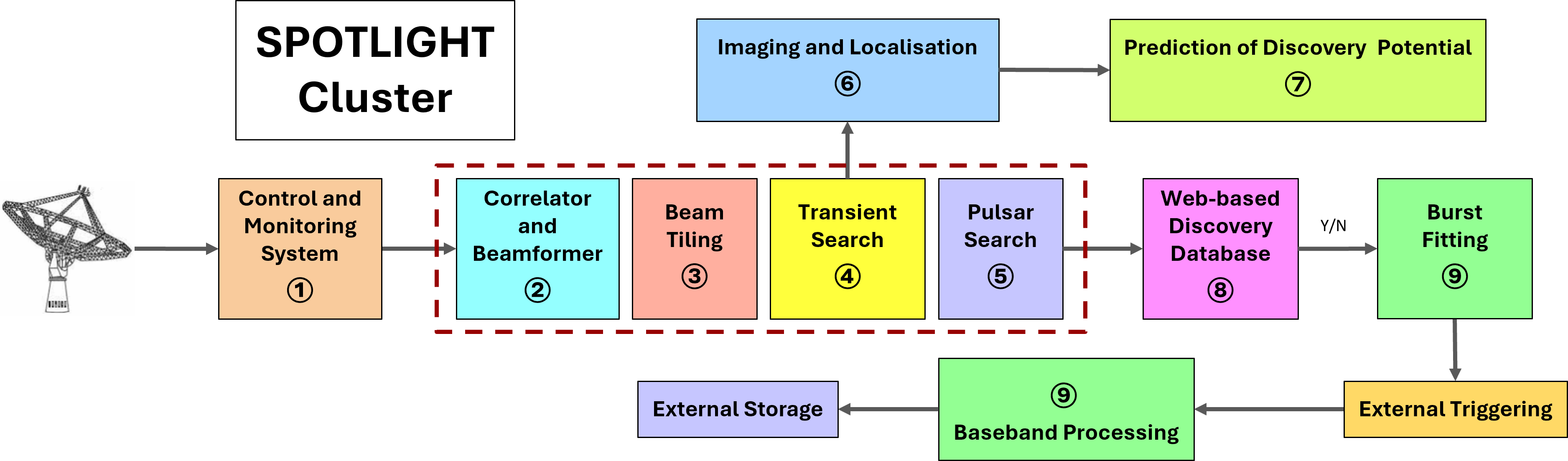}
            \caption{Block Diagram of the SPOTLIGHT HPC cluster showing all the technical components.}
            \label{fig:spotlight_block}
        \end{figure}

\begin{figure}%
            \centering
            \subfloat[]{\includegraphics[width=7cm]{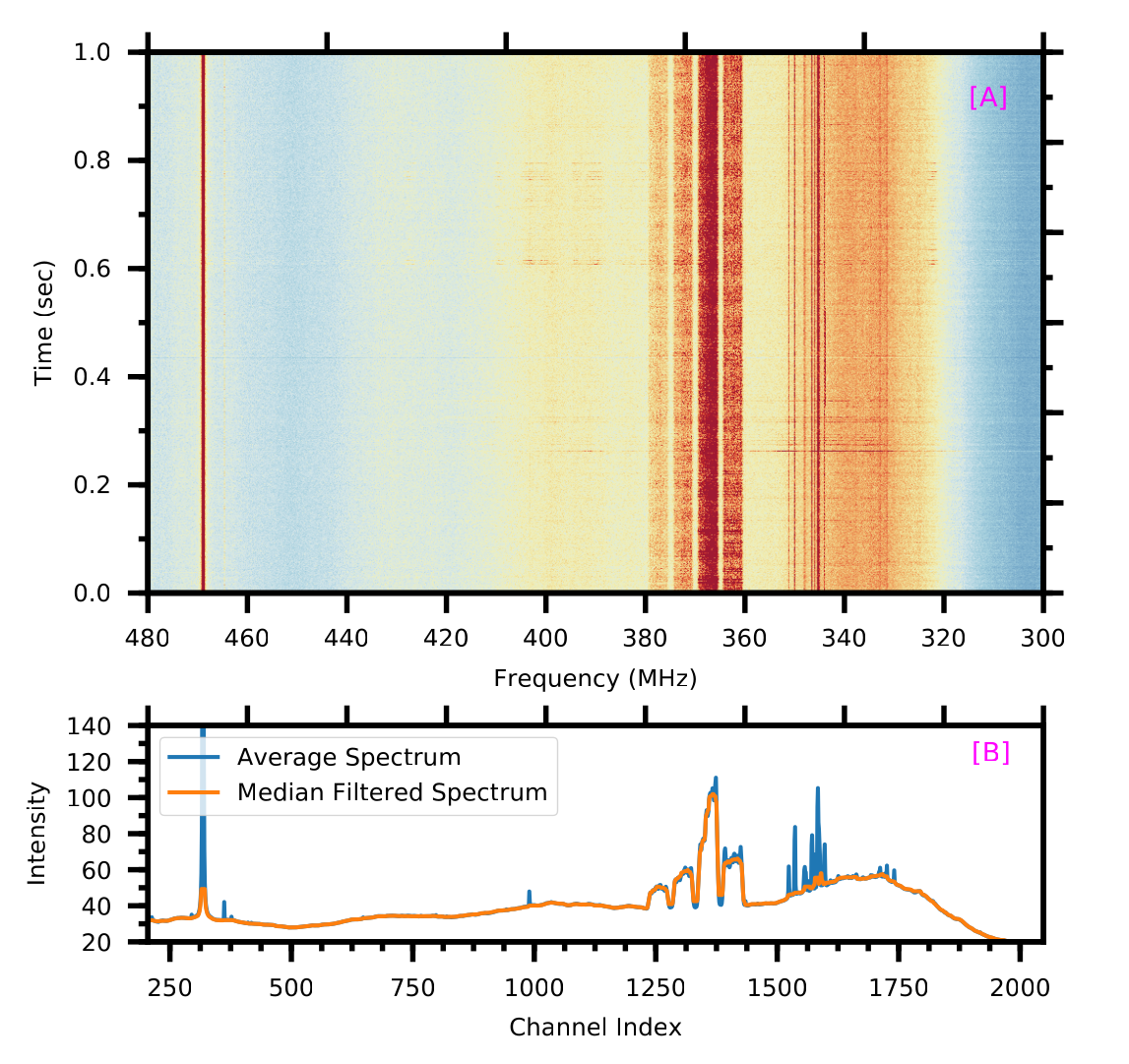}}\\%
            \subfloat[]{\includegraphics[width=7cm]{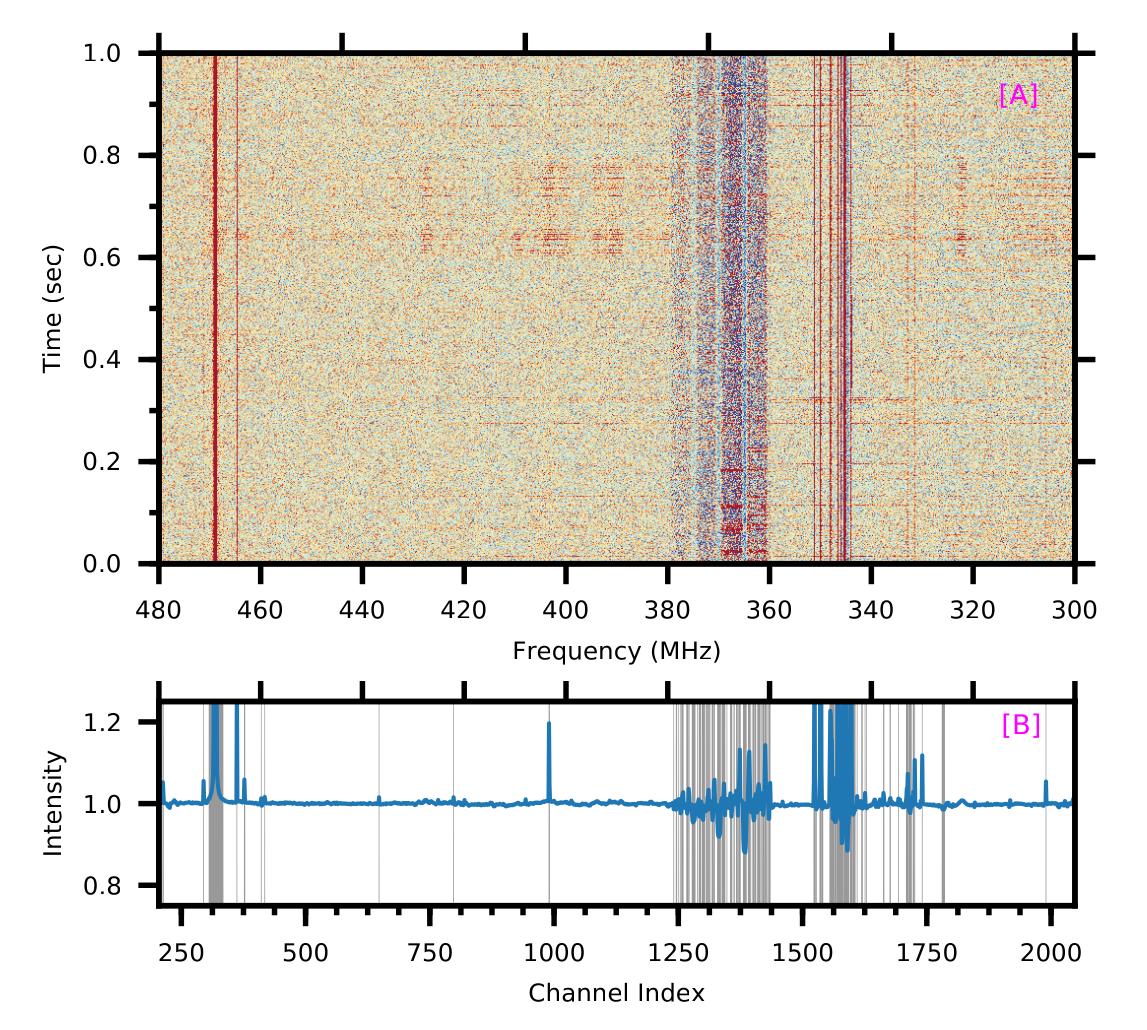}}%
            \qquad
            \subfloat[]{\includegraphics[width=7cm]{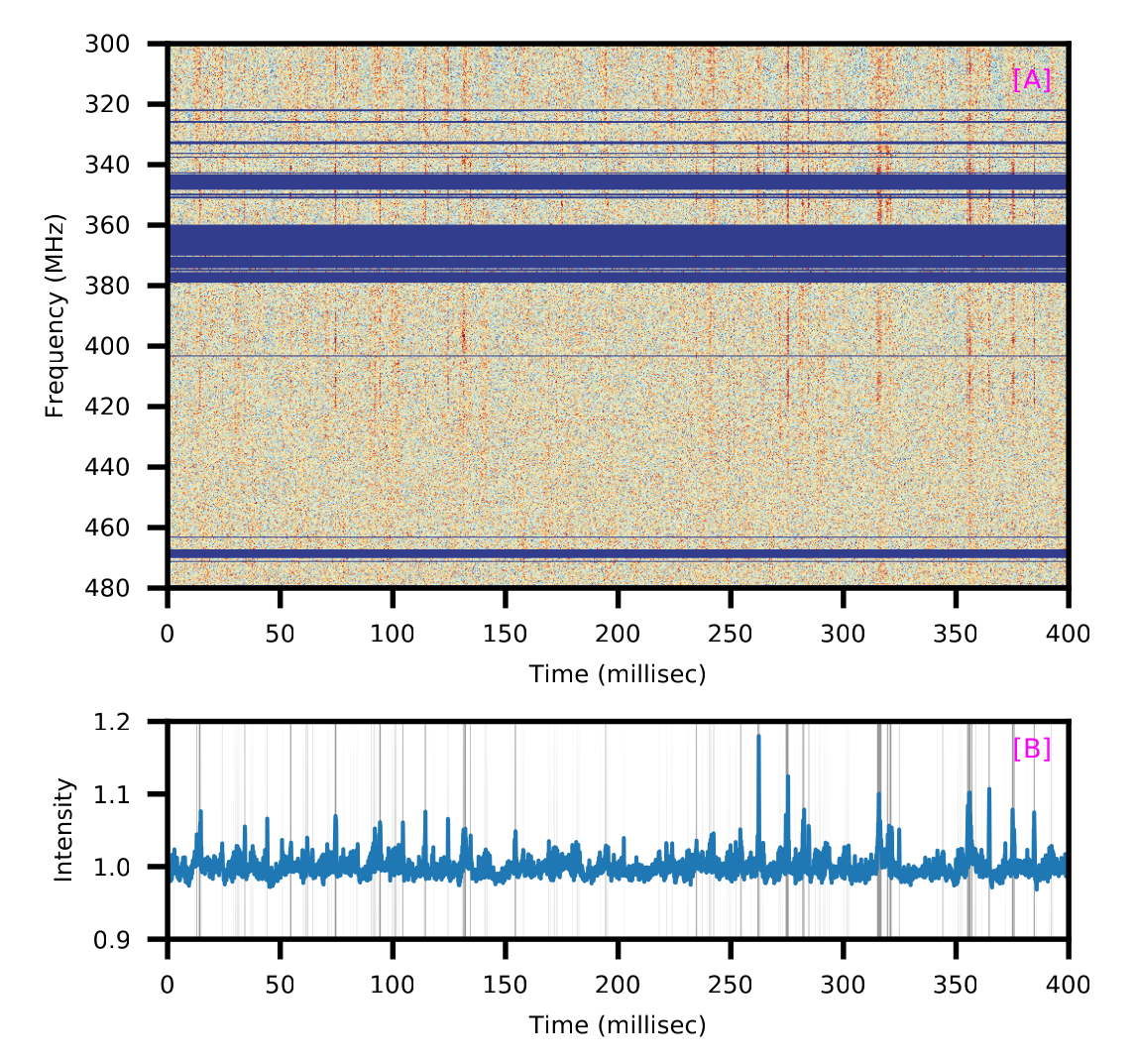}}%
            \caption{(a) [A]: Example of beamformer data recorded using the uGMRT Band-3 (250 MHz-500 MHz) receivers; intensity is plotted in colour as a function of observing time and spectral channel. [B] The average spectrum obtained by taking the mean of the intensity for each spectral channel (blue curve) and the median filtered spectrum obtained using a moving median window of width 21 channels (orange curve). (b) [A]: The same data as displayed in figure 3 after correcting for the system bandpass; intensity is plotted in colour as a function of observing time and spectral channel. [B] The blue curve shows the average spectrum of the data in [A], obtained by taking the mean of intensity in each spectral channel. The channels which are detected as > 4.0$\sigma$ outliers in the spectrum are shaded in grey. (c) [A]: The first 500 ms of the bandpass-corrected data that is displayed in figure 4[A]; intensity is plotted in colourway r as a function of observing time and spectral channel. All detected spectral-line RFI have been flagged; these channels are marked in dark blue. [B] The blue curve shows the frequency-averaged time series of the data in [A], obtained by taking the mean of intensity in each time sample; flagged spectral channels are ignored while taking the average. The time samples which are detected as > 3.0$\sigma$ outliers in the time-series are shaded in grey. Credit for all the plots: Aditya Chowdhury and Yashwant Gupta}%
            \label{fig:gptool}%
        \end{figure}

\begin{figure}
            \centering
            \includegraphics[width=15.5cm]{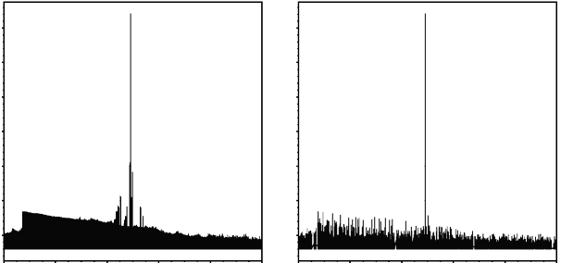}
            \caption{This figure shows the effect of peaks filtering on the number of candidates. The $x$-axis is time and the $y$-axis is SNR. Each candidate is represented by a vertical line. Credits: Ujjwal Panda}
            \label{fig:peaks_filtering}
        \end{figure}

\begin{figure}
            \centering
            \includegraphics[width=10cm]{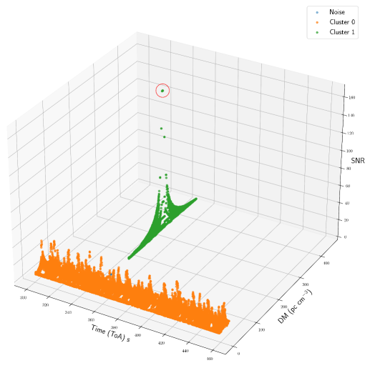}
            \caption{Candidates output from the \href{https://github.com/AstroAccelerateOrg/astro-accelerate}{AstroAccelerate} (\cite{armour_astroaccelerate_2020}, \cite{novotny_accelerating_2023}, \cite{adamek_single-pulse_2020}) software package clustered by the clustering DBSCAN-based algorithm (\cite{ester_density-based_1996}). The candidate with the highest SNR in the cluster of the real signal (green) is circled in red. Candidates coloured in orange are interferences. Credit: Ujjwal Panda}
            \label{fig:clustering}
        \end{figure}

\begin{figure}
            \centering
            \includegraphics[width=10cm]{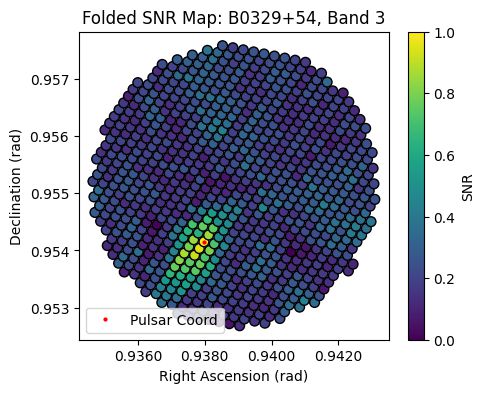}
            \caption{The sky SNR map of the folded profile of the pulsar B0329+54 observed on January 01, 2025 in the band 3 of uGMRT. Each circle represents one beam. The pulsar coordinates are marked with a red dot. As is apparent, the pulsar can be seen in many nearby beams appearing like a cluster. We expect a similar sky SNR map for a single pulse FRB. Credit: Raghav Wani}
            \label{fig:snr_map}
        \end{figure}

\begin{figure}
            \centering
            \includegraphics[width=10cm]{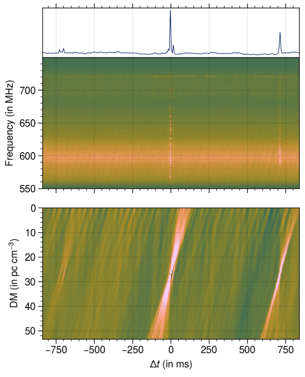}
            \caption{A high-SNR candidate whose features have been extracted using \href{https://github.com/astrogewgaw/candies}{candies}. The top panel is the dedispersed time series. The middle panel is the dedispersed dynamic spectrum. The bottom panel is the DM transform, a.k.a. the ``bow-tie diagram". Credits: Ujjwal Panda}
            \label{fig:candies}
        \end{figure}

\begin{figure}
            \centering
            \includegraphics[width=10cm]{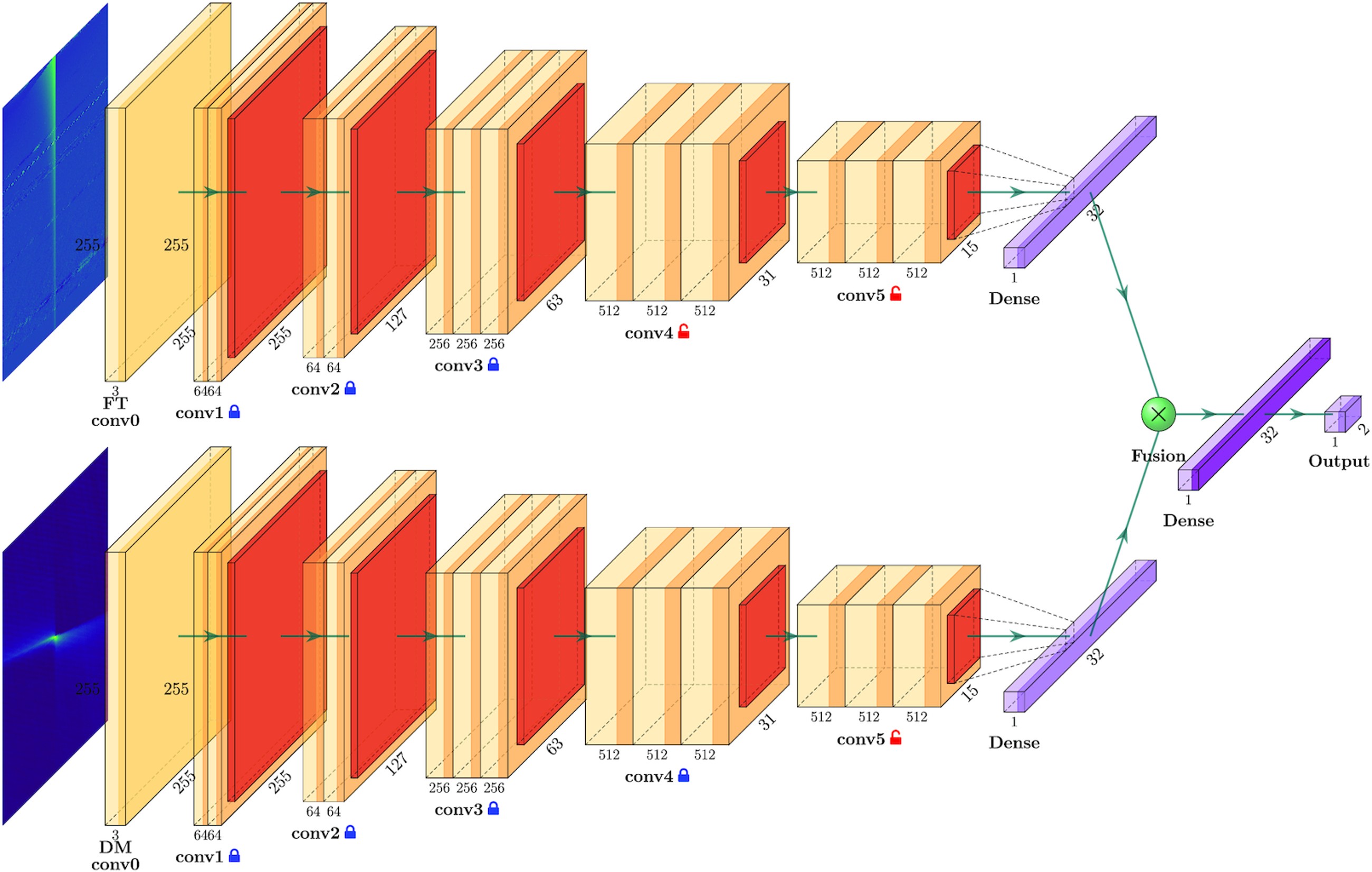}
            \caption{A sample network architecture available in \href{https://github.com/devanshkv/fetch}{FETCH}. (\cite{agarwal_fetch_2020})}
            \label{fig:fetch}
        \end{figure}

        The SPOTLIGHT cluster (shown in \figref{fig:spotlight_block}) is a multi-component HPC system\footnote{The Param Rudra, funded under the \href{https://www.nsmindia.in/}{National Super Computing Mission (NSM) of India}, built and maintained by the \href{https://www.cdac.in/}{Centre for Development of Advanced Computing (C-DAC), Pune}} consists of a telescope control and monitoring system, a multi-beam correlator and beamformer software, a beam synthesis and tiling software, a real-time multi-beam transient search pipeline, a quasi-real-time multi-beam pulsar search pipeline, an imaging and localisation pipeline, an interface to predict the discovery potential of the candidate events, a web-based discovery database, a burst fitting software, an external triggering system to trigger the recording of the baseband (the full Nyquist-resolution) data and further processing like measuring the polarisation. The transient search pipeline processes the multi-beam data output by the correlator and beamformer software to detect dispersed single pulses in an automated workflow. It consists of a radio frequency interference (RFI) mitigation software, a module each for de-dispersion, single pulse search (SPS) and peaks filtering integrated into the \href{https://github.com/AstroAccelerateOrg/astro-accelerate}{AstroAccelerate} (\cite{armour_astroaccelerate_2020}, \cite{novotny_accelerating_2023}, \cite{adamek_single-pulse_2020}) software package, a clustering program based on the Density-Based Spatial Clustering of Applications with Noise (DBSCAN) algorithm (\cite{ester_density-based_1996}), a feature extraction software (\href{https://github.com/astrogewgaw/candies}{candies}) for FRBs and a CNN-based binary classifier (\href{https://github.com/devanshkv/fetch}{FETCH}). Figures \ref{fig:gptool}-\ref{fig:fetch} highlight the data processing workflow by showing the intermediate data products.
        
        \begin{figure}
            \centering
            \includegraphics[width=15.5cm]{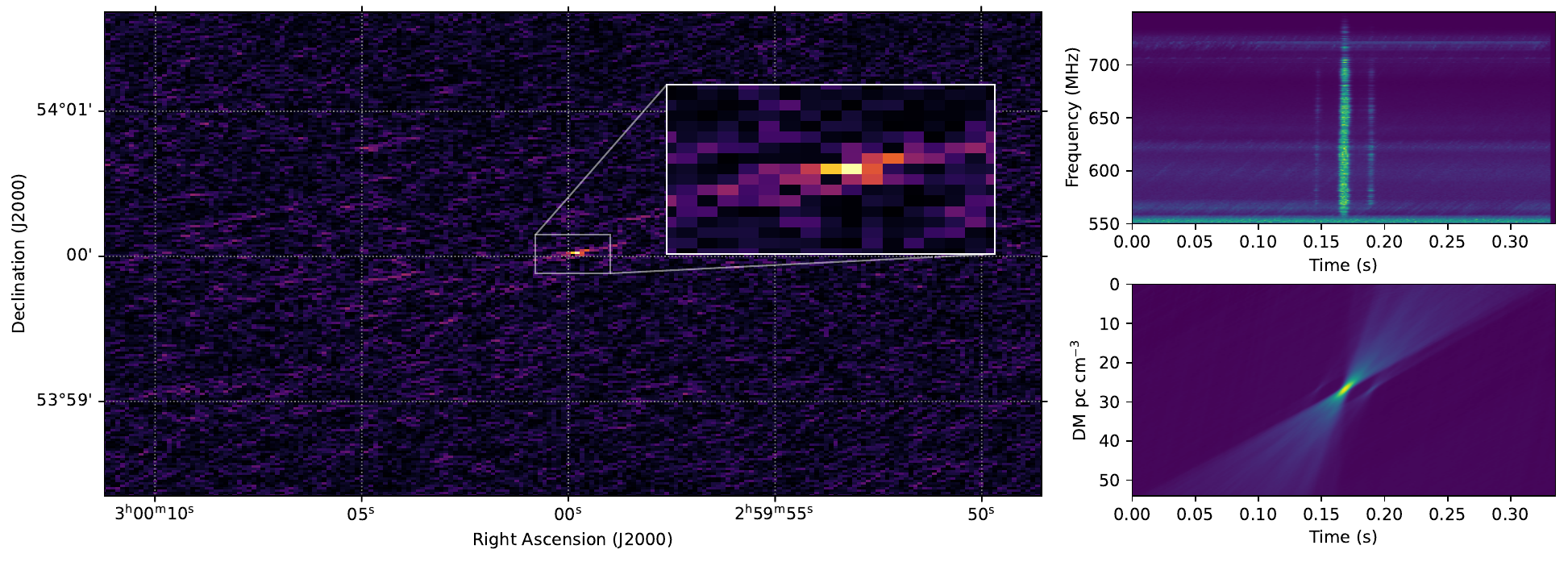}
            \caption{Left: A single pulse from a pulsar localised by imaging the $\sim$1.3 ms visibilities. Note: The RA and Dec. are off from their expected values due to some internal processing error. Top right: Dedispersed dynamic spectrum of the same single pulse. of the same single pulse. Bottom right: the DM transform, a.k.a. the ``bow-tie diagram", of the same single pulse. Credits: Arpan Pal}
            \label{fig:localisation}
        \end{figure}
        
        The RFI mitigation software, \href{https://github.com/chowdhuryaditya/gptool}{GMRT Pulsar Tool (gptool)}, does bandpass correction and iteratively does both, time- and frequency-domain flagging twice (see\figref{fig:gptool}). The de-dispersion module de-disperses the beam data for a large number of trial DMs from $0 - 2000\ \rm{cm^{-3}kpc}$.  The SPS module searches for statistically significant single pulses above a user-specified threshold using time-domain boxcar filtering. The list of candidates thus generated is filtered to remove redundant candidates from a singular event using square stencils of different sizes in the DM-time plane (see \figref{fig:peaks_filtering}). The clustering algorithm forms clusters of candidates in the DM-time plane and picks out the highest SNR candidates from each cluster(see \figref{fig:clustering}). Clusters of these candidates in the RA-Dec plane (see \figref{fig:snr_map}) are identified and the candidates with the highest SNR in their respective clusters are retained; the rest are discarded. The peaks filtering and clustering steps reduce the number of false and redundant candidates significantly. The feature extraction software, \href{https://github.com/astrogewgaw/candies}{candies}, forms the characteristic de-dispersed dynamic spectrum and the DM-transform (the bow-tie diagram) of the retained candidates (see \figref{fig:candies}). The CNN-based classifier, \href{https://github.com/devanshkv/fetch}{FETCH} (\cite{agarwal_fetch_2020}), classifies the candidate as either an interference or a true dispersed signal based on a confidence score threshold (see \figref{fig:fetch}). The baseband data corresponding to the positively classified candidates is dumped to the external Petabyte storage. Parallelly, the imaging and localisation pipeline makes images of the positively classified candidates from the $\sim$1.3 ms visibilities, localises the signal in the sky and estimates its fluence (see \figref{fig:localisation}). The final data products are stored in the web-based discovery database and are made available for evaluating the discovery potential by human scrutiny. The data is also made available to the user for burst analysis and baseband voltages processing at full time- and frequency-resolution.
        
    \section{Interfacing the transient search pipeline with the beamformer}\label{sec2.2}
        The fine time-resolution required to search for short-duration transients makes it unfeasible to store the dynamic spectrum for offline processing. Thus arises the need to buffer the huge volume of the multi-beam dynamic spectrum in the system memory for real-time searching. This is achieved by utilising a significant amount of host memory (1200 MB per beam\footnote{1 MB = $1024^2$ bytes.}, to be precise) to maintain a ring buffer to retain $\sim$402.65 s of the dynamic spectrum at any given time. The buffer has to be a shared memory (SHM) construct to enable seamless parallel access to multiple processes running simultaneously, in parallel to each other. The inter-process communication (IPC) shared memory segments provide a fast and reliable construct to meet this requirement.
        
        \begin{figure}
            \centering
            \includegraphics[width=15.5cm]{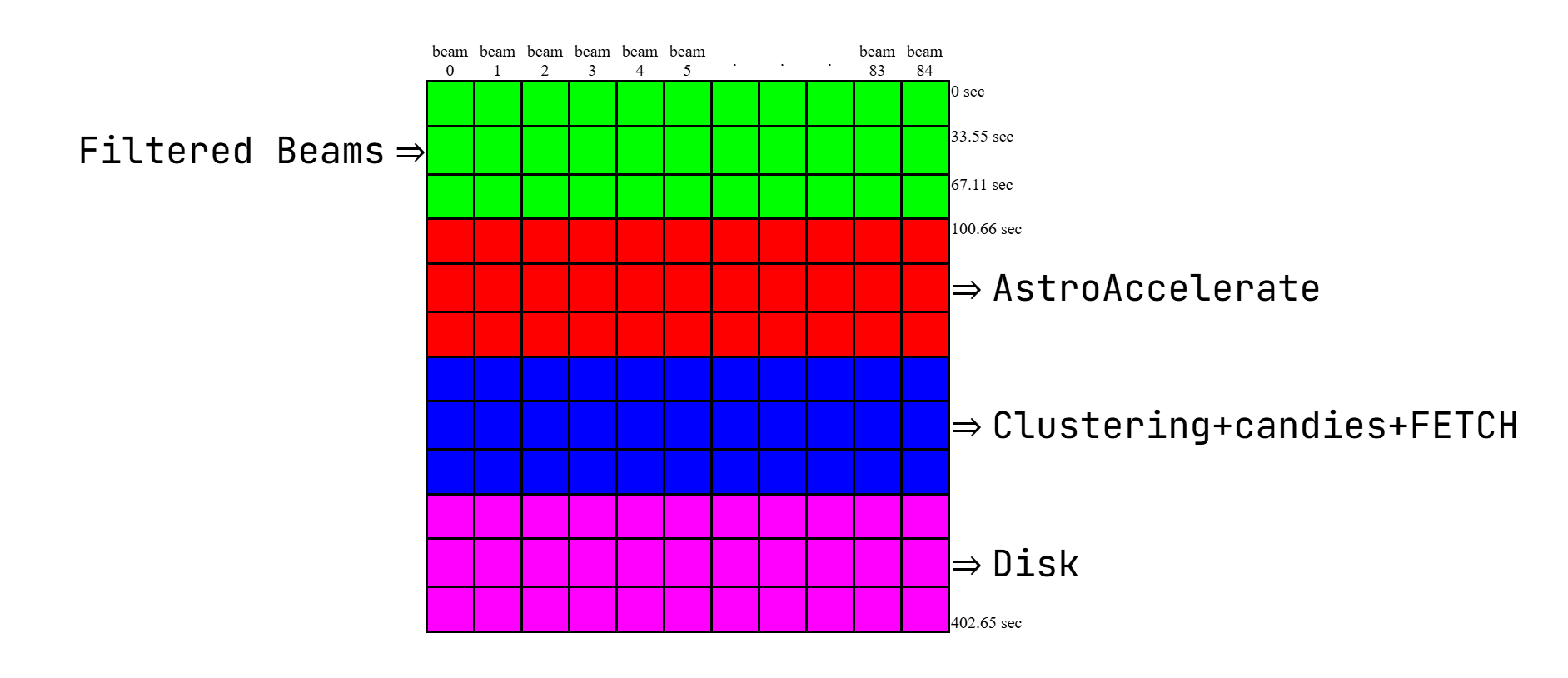}
            \caption{The 2-D grid structure of the FRB\_SHM. In this schematic, the number of beams shown is 85; each row represents a block with the squares representing the beams.}
            \label{fig:frb_shm}
        \end{figure}
        
        The SPOTLIGHT HPC cluster consists of 60 compute nodes and 2 login nodes. The SPOTLIGHT backend (correlator and beamformer) runs on 16 of the compute nodes constituting node-group 1 and will form $\sim$2000 beams\footnote{It currently forms 800 beams.}. The transient search pipeline will run on 24 of the compute nodes\footnote{It currently runs on 16 of the compute nodes.} constituting node-group 2. The beamformed dynamic spectrum is transported from node-group 1 to 2 via a message passing interface (MPI) and is buffered in a smaller (25 MB per beam or $\sim$8 s long) IPC SHM construct (referred to as TEL\_SHM henceforth). Each node will thus host $\sim$85 beams\footnote{Currently, each node hosts 50 beams.}. The TEL\_SHM is divided into 8 blocks each consisting of 800 spectra of length 4096 frequency channels from all the beams. A 2-D grid of beams and blocks is thus formed by 2-D indexing of a linear memory segment. The TEL\_SHM is accumulated in a 48 times larger FRB\_SHM. The FRB\_SHM is divided into 12 blocks each consisting of 25600 spectra of 800 spectra of length 4096 frequency channels from all the beams, forming a similar 2-D grid as the TEL\_SHM (see \figref{fig:frb_shm}). This (large) SHM buffer is maintained in real-time for \href{https://github.com/AstroAccelerateOrg/astro-accelerate}{AstroAccelerate} and candies software of the transient search pipeline to read and process the beam data. The depth (temporal extent) of the FRB\_SHM ensures that the cumulative pipeline delay in the real-time processing is appropriately accounted for. All the observation and beam specific metadata is buffered in separate IPC SHM segments known as FRB\_SHM headers on each host node along with the FRB\_SHM. Both, the FRB\_SHM headers and the FRB\_SHM body segments, are of user-defined C/C++ structured data types borrowed from the GWB beamformer and modified to suit the needs of the SPOTLIGHT backend.
    
    \section{Towards incorporating an RFI mitigation software in the SHM interface}\label{sec2.3}
        The beamformed dynamic spectrum output by the backend is often contaminated by RFI. Although the local (affecting only a few antennae) RFI gets ``washed-out" significantly because of discarding the antennae self terms in the PCPA beams, strong and non-local RFI can greatly reduce the survey sensitivity. The \href{https://github.com/chowdhuryaditya/gptool}{GMRT Pulsar Tool (gptool)} has been widely used for RFI mitigation and pulsar processing of a single beam from GMRT's GWB beamformer data. It iteratively filters time- and frequency-domain RFI (see \figref{fig:gptool}). I spent a great deal of time in understanding the intricate workings of the software with the aim of incorporating the RFI mitigation functionality into the SHM interface between the TEL\_SHM and the FRB\_SHM. The latest single-beam version of gptool was modified by its author to bypass the pulsar processing tasks. It was capable of reading GMRT's GWB beamformer data and write the filtered data into an older version of the larger FRB\_SHM. The approach taken was to start from the full complexity code and remove the unwanted functionality of visualisation and pulsar processing. The end result was expected to be a software only for multi-beam RFI mitigation, which reads data from the multi-beam TEL\_SHM, filters the RFI from all the beams in a combination of sequential and parallel processing on multiple threads and writes the filtered data to the multi-beam FRB\_SHM. This task, however, couldn't be completed within the time constraints necessary to meet the SPOTLIGHT project development expectations.
    
    \section{Testing the transient search pipeline}\label{sec2.4}
        With a large collaboration sharing a common development goal and the leaps of progress it made possible, there was an incessant need to test multiple components of the end-to-end transient search pipeline with the limited telescope time available during the maintenance period. Especially before the inauguration of the Param Rudra system, limited testing time and missing links in the search pipeline made it necessary to have a software to inject simulated FRBs with known pulse parameters. It also enables us to test the completeness of our survey with respect to the FRB parameter space.
    
        An ensemble of FRB single pulses with known parameters like DM, radio frequency, pulse geometry, width and flux, etc. were simulated using the \href{https://bitbucket.csiro.au/scm/psrsoft/simulatesearch.git}{SIMULATESEARCH} (\cite{luo_simulating_2022}) software. It provides a module to inject these simulated single pulses in noise data. The current version of the software, however, has the limitation of being able to inject pulses in noise data of up to 4-bits digitisation. Thus, it cannot be used to inject pulses in the 8-bit digitised data of GMRT.
        
\begin{figure}
            \centering
            \includegraphics[width=14cm]{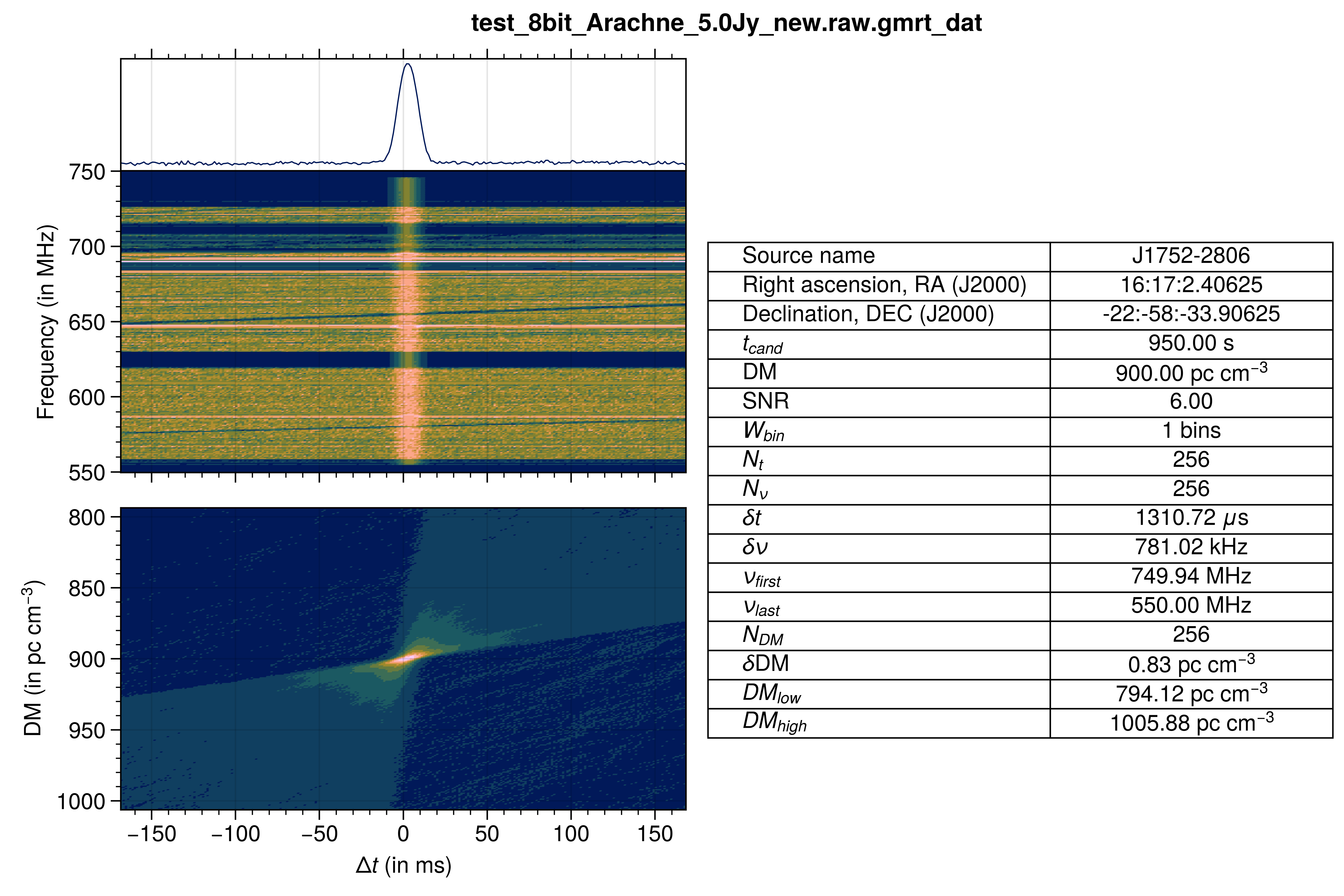}
            \caption{A simulated FRB with a Gaussian pulse profile injected into real noise data using \href{https://github.com/astrogewgaw/arachne}{arachne}, detected using the \href{https://github.com/AstroAccelerateOrg/astro-accelerate}{AstroAccelerate}-based pipeline and featurised using \href{https://github.com/astrogewgaw/candies}{candies}. The left panel is the same as \figref{fig:candies}. the right panel shows the observation details and candidate parameters. Credits: Kenil Ajudiya, Ritavash Debnath, Raghav Wani and Ujjwal Panda}
            \label{fig:arachne}
        \end{figure}

        Since most of the values of the dynamic spectrum of a simulated pulse (without any noise) are zeros, the simulated pulses are converted to a sparse matrix binary file format (tuples containing time bin, frequency channel and value). This not only saves storage space, but also reduces injection time because of a significant reduction in the number of iterations required. The \href{https://github.com/astrogewgaw/arachne}{arachne} software reads simulated pulses from the sparse matrix files and weaves in fake FRBs in live GMRT data from the TEL\_SHM. Injection into 8-bit digitised data is achieved by extending the logical formulation presented in \cite{luo_simulating_2022} for 1-bit and 2-bit digitisation to 8-bit digitisation. I contributed in working out the logical formulation and in the debugging of the code written by a colleague. \figref{fig:arachne} shows an example single pulse injected using \href{https://github.com/astrogewgaw/arachne}{arachne}, detected using the \href{https://github.com/AstroAccelerateOrg/astro-accelerate}{AstroAccelerate}-based pipeline and featurised using \href{https://github.com/astrogewgaw/candies}{candies}.
        
        Before the multi-beam SPOTLIGHT backend was available, the tests used the single-beam version of gptool which was modified to bypass the pulsar processing tasks. Simulated FRBs were then injected into the filtered data and buffered into a structural copy of the FRB\_SHM. The \href{https://github.com/AstroAccelerateOrg/astro-accelerate}{AstroAccelerate} software attached to the FRB\_SHM having the injected FRBs and outputs candidate signals which are then clustered in the DM-time plane as described in \secref{sec2.1}. The highest SNR candidate of each of the clusters were directly classified using FETCH (skipping the RA-Dec. clustering step until its implementation as an additional coincidence-anti-coincidence clustering algorithm). These tests were primarily focused around functionality testing and debugging of all the pipeline components. The rigorous planning, execution and documentation of these tests enabled a speedy development of the real-time multi-beam transient search pipeline it currently is, at a stage very close to final deployment in a commensal mode.
    
    \chapter{Developing the Pulsar Search Pipeline}\label{ch3}
    \section{Overview of the pulsar search pipeline}\label{sec3.1}
        Since pulsars have a well-defined periodic emission, a long duration observation can be used to detect it at high statistical significance even for the faint pulsars. It is therefore, beneficial to store the dynamic spectrum of the entire duration of the observation rather than a real-time buffering system which can retain only reasonably short duration of observations. The storage space limitations still necessitates the analysis of the stored data in a quasi-real-time fashion while the observatory undergoes weekly maintenance. Even then, not all of the $\sim$2000 beams can be stored in the Petabyte storage available to SPOTLIGHT. Thus, a basic, quick periodicity search based on \href{https://github.com/jack-white1/pulscan}{Pulscan} (\cite{white_pulscan_2024}) is aimed to be deployed for real-time periodicity search across all the beams. A broader, Fourier Domain Acceleration Search (FDAS), will be conducted offline in quasi-real-time using \href{https://github.com/AstroAccelerateOrg/astro-accelerate}{AstroAccelerate} (\cite{armour_astroaccelerate_2020}). Both the searches will report candidate pulsar signals. The data then needs to be folded according to the candidate's period and period derivatives to obtain a folded profile as a function of time and frequency which will be passed on to a CNN-based binary classifier, \href{https://github.com/scienceguyrob/GHVFDT}{Gaussian Hellinger Very Fast Decision Tree (GH-VFDT)} (\cite{lyon_tree_2014}), for a confidence score-based classification. Some of the candidate parameters like its DM, period and period derivatives obtained from the searches can be further optimised for highest possible signal-to-noise ratio (SNR) of the folded profile. The parameter optimisation requires a relatively narrower search around the tentative parameters, but since the search is fine and multi-dimensional, it is computationally expensive. The large data volumes further necessitate GPU acceleration of the folding-based search algorithm. Moreover, such a ``brute-force" search algorithm accelerated on modern GPUs can also be deployed for a direct search (without FDAS or FDJS) to overcome the limitations of the lesser compute expensive Fourier domain techniques.
        
        \begin{figure}%
            \centering
            \subfloat[]{\includegraphics[width=6cm]{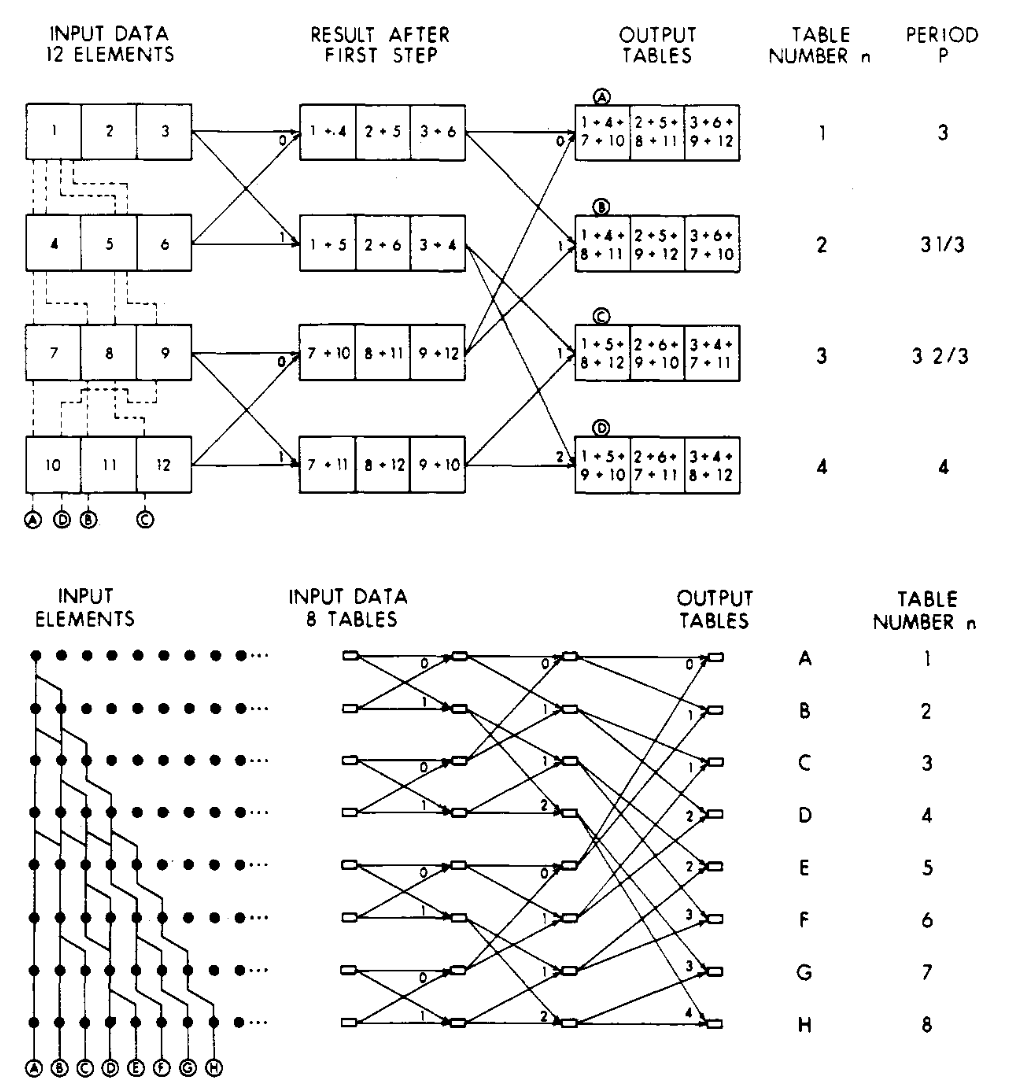}}%
            \qquad
            \subfloat[]{\includegraphics[width=8.5cm]{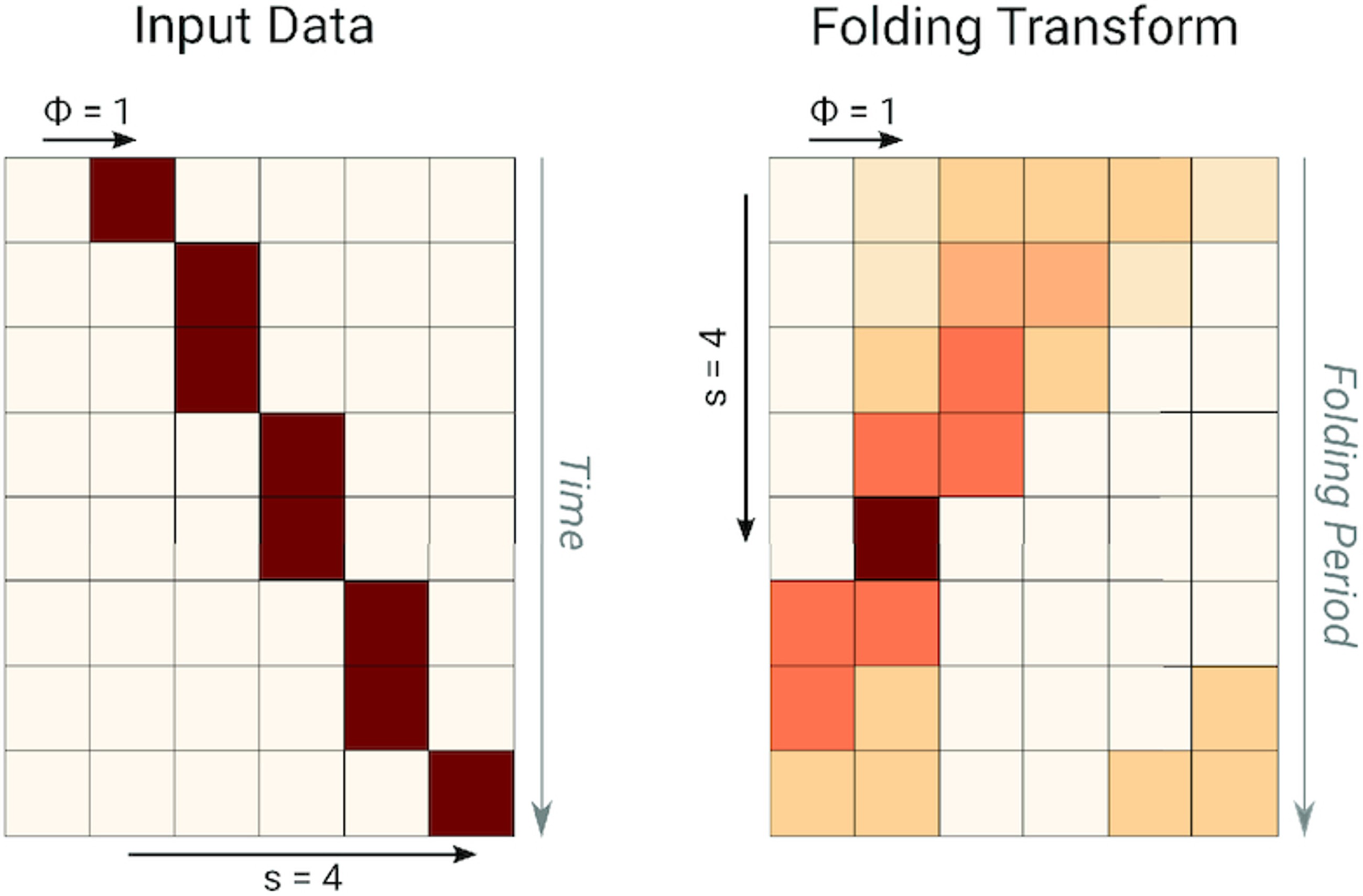}}%
            \caption{Schematic representation of the fast folding algorithm. (a) Source: \cite{staelin_fast_1969}. (b) Source: \cite{morello_optimal_2020}}%
            \label{fig:FFA}%
        \end{figure}
        
        The Fast Folding Algorithm (FFA) (\cite{staelin_fast_1969}) is a phase-coherent search technique for periodic signals. It directly and efficiently folds the data at a wide range of narrowly spaced trial periods by taking advantage of redundant operations (see \figref{fig:FFA} for an overview of the algorithm). A folded profile is produced for each trial period, which can then be evaluated for the presence of a pulse. Despite of the research advances in scientific computation, to the best of our knowledge, a high throughput, GPU-accelerated implementation of an FFA-based pulsar search is not available. It has several advantages over the Fourier domain methods, in that it closely approaches the theoretical optimum sensitivity to all periodic signals and it is analytically shown to be significantly more sensitive than the standard fast Fourier transform (FFT) with incoherent harmonic summing (IHS) method, regardless of pulse period and duty cycle (\cite{morello_optimal_2020}). However, it is limited to be able to fold strictly periodic signals only. This is a prohibitive issue since a significant fraction of all the known pulsars are found in binary systems - their signals are no longer strictly periodic. In order to overcome this limitation, we correct for the pulsar's period derivatives (up to the second derivative since higher derivatives imply more search dimensions) by re-sampling the time series at a time-dependant sampling frequency before folding. This should not be mistaken with transforming the observed time series to the emitter's rest frame because the pulsar signal can have inherent period derivatives because of the spin-down of its emitter. The following sections describe the processing workflow starting from the dynamic spectrum.
        
    \section{Dedispersing the dynamic spectrum}\label{sec3.2}
\begin{figure}
            \centering
            \includegraphics[width=15.5cm]{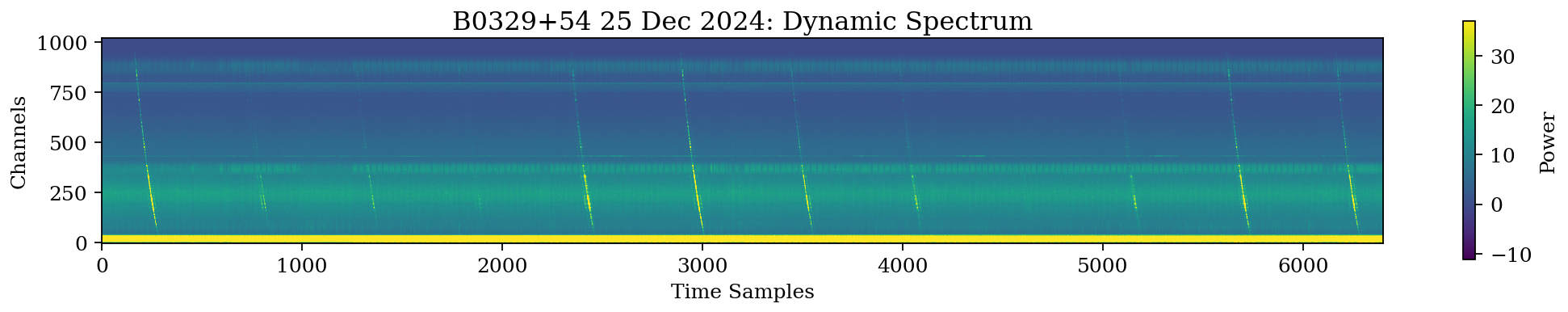}
            \caption{The dynamic spectrum of the pulsar B0329+54 observed on December 25, 2024 in the band 4 of uGMRT. Bright single pulses are clearly visible with their characteristic dispersion sweep. The data is not corrected for the bandpass.}
            \label{fig:dynamic_spectrum}
        \end{figure}

\begin{figure}%
            \centering
            \subfloat[]{\includegraphics[width=15.5cm]{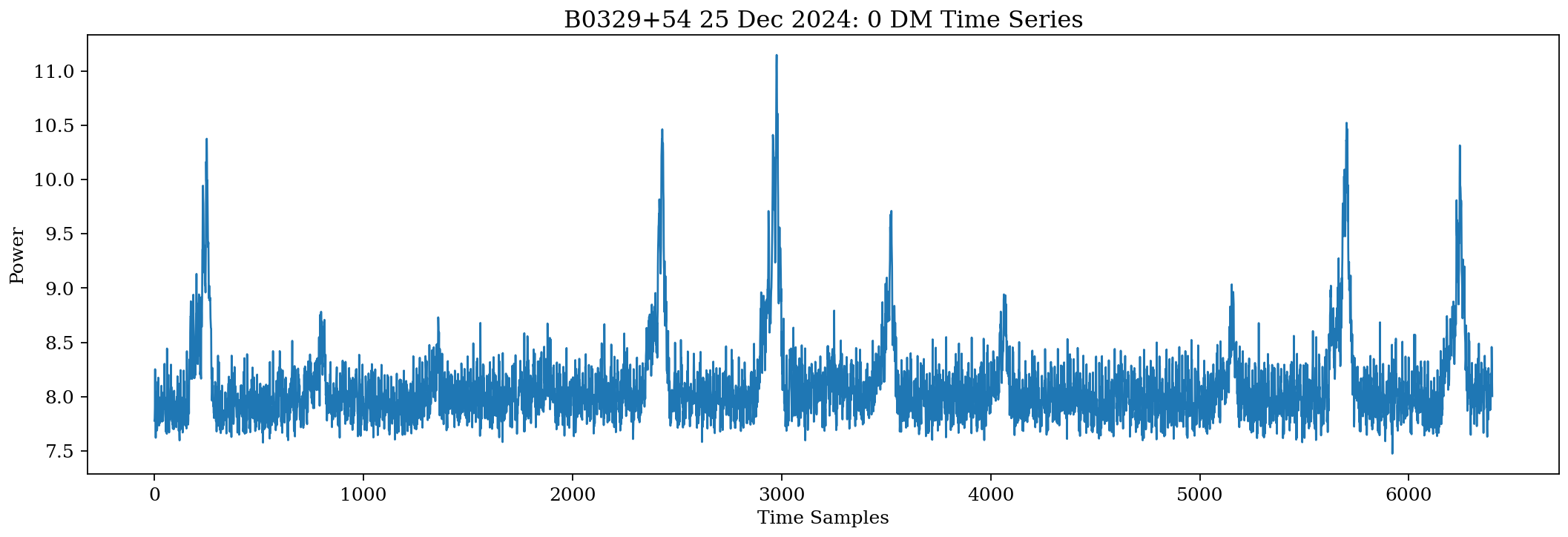}}\\%
            \subfloat[]{\includegraphics[width=15.5cm]{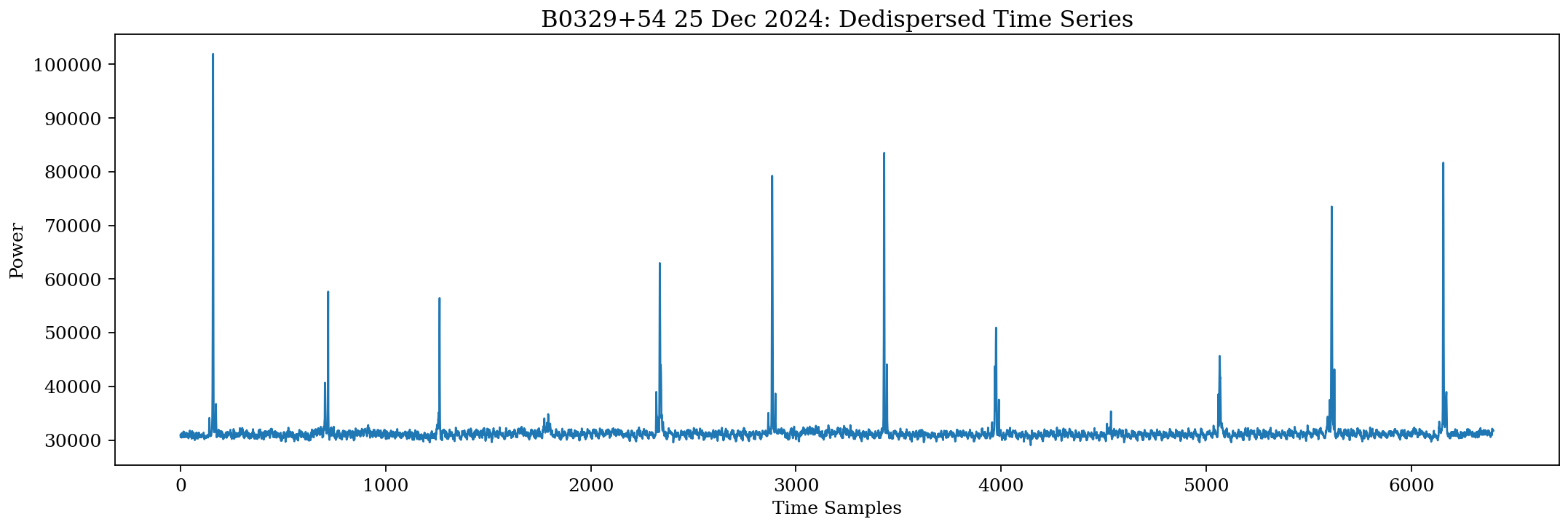}}%
            \caption{(a) Time series of the dynamic spectrum obtained without correcting for the dispersion delay, a.k.a. the ``0 DM" time series. The peaks get smeared into multiple time bins as a result of dispersion, resulting in SNR loss. (b) The dedispersed time series with sharp peaks indicating a small duty-cycle.}%
            \label{fig:dedispersion}%
        \end{figure}

        As mentioned in \chref{ch1}, the pulsar signal carries a characteristic dispersion sweep (see \figref{fig:dynamic_spectrum}), which, if not accounted for, results in the smearing of the pulsed emission in multiple time bins in the time series (see \figref{fig:dedispersion}) and subsequent SNR loss. The brute-force dedispersion is an $O(m \times N)$ algorithm where $m$ is the number of frequency channels and $N$ is the number of time bins in the dynamic spectrum. Hence, it is a very slow and compute expensive processing step. When compared with all the other steps described below, dedispersion was indeed the slowest step by an order of magnitude or even more in terms of execution time. A simple parallelisation when implemented in CUDA/C++ and executed on NVIDIA GeForce RTX 3090 GPU reduced the execution time by about 6 times compared to the single thread version executed on an AMD Ryzen CPU core (improvement from a real-time factor\footnote{Real-time factor of an algorithm = length of the observation / execution (or processing) time} of $\sim$100 to $\sim$600 on an average). The dedispersed time series is then downsampled by a user-input factor is necessary.
        
    \section{De-reddening the times series}\label{sec3.3}
\begin{figure}%
            \centering
            \subfloat[]{\includegraphics[width=15.5cm]{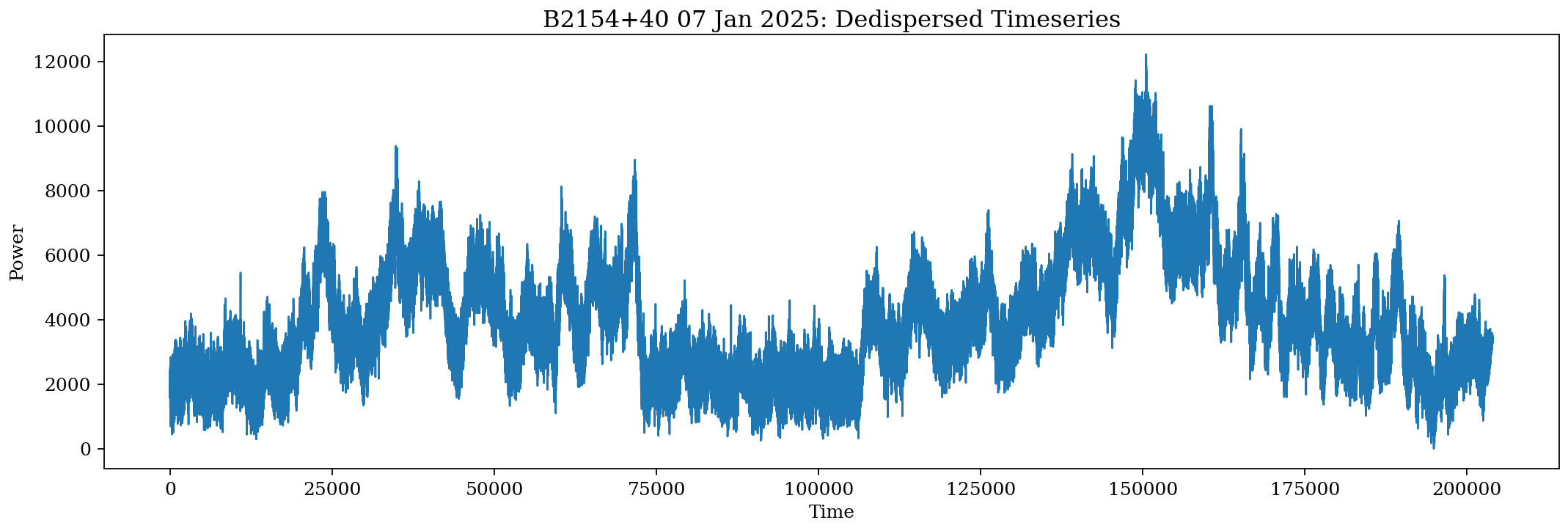}}\\%
            \subfloat[]{\includegraphics[width=15.5cm]{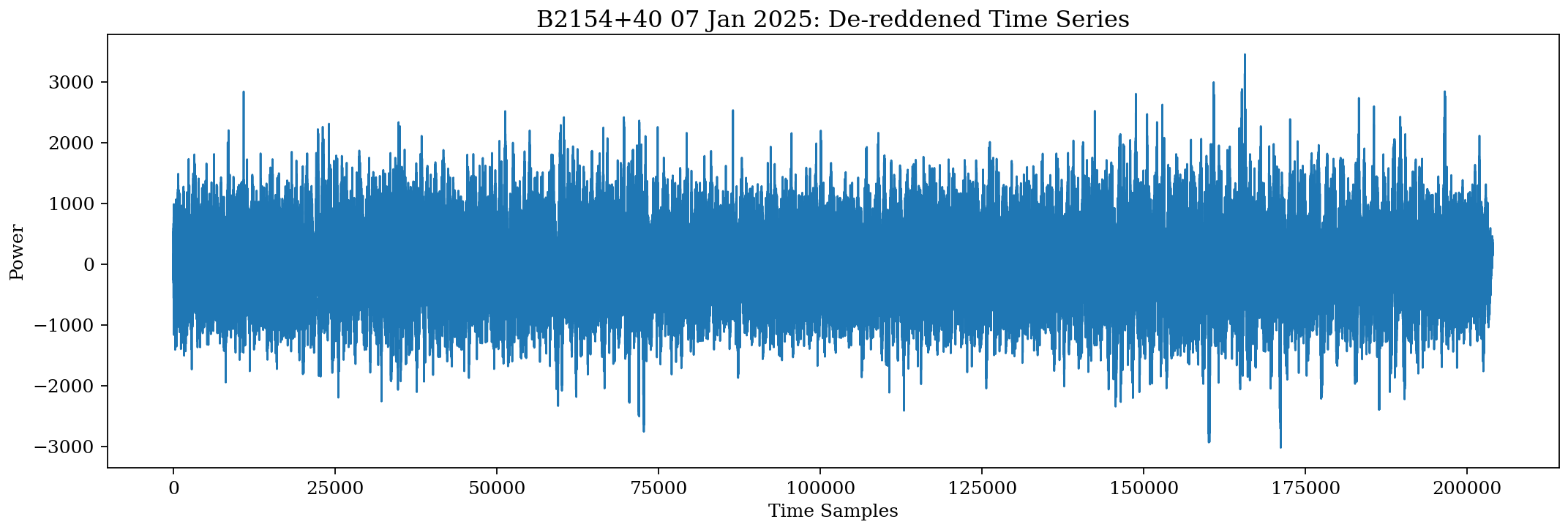}}%
            \caption{(a) The dedispersed time series of a faint pulsar B2154+40 observed on January 07, 2025 in the band 3 of uGMRT. The baseline fluctuations are due to heavy contamination from RFI. (b) The above time series de-reddened using a running window median subtraction algorithm. The fluctuations are reduced dramatically, but the pulsar signal is still buried under the noise since no RFI mitigation tool was used during the observation.}%
            \label{fig:median_sub}%
        \end{figure}

        The time series of long duration observations often have a fluctuating baseline primarily because of antenna gain fluctuations and periodic RFI like those from AC power cables. This, slow-varying baseline noise constitutes the so-called ``red-noise". It has been shown that the de-reddening of the time series using running-median subtraction is more sensitive to long-period pulsars than the spectral whitening technique (see, for example, \cite{singh_gmrt_2022}, \cite{van_heerden_framework_2017} and \cite{lazarus_arecibo_2015}). Moreover, median is less sensitive to outliers than the mean as an estimate of the noise in pulsar time series. Running median subtraction was thus preferred over other de-reddening methods. \figref{fig:median_sub} shows the action of running median subtraction on a fluctuating time series. One of the challenges here is to determine the window size to be used. It should be less than half of the timescale of the small fluctuations one wishes to eliminate and must be more than twice the maximum expected pulse width. For the purposes of SPOTLIGHT operations, this shall be fixed at a later stage after rigorous testing on real noise and signal data.
        
    \section{Re-sampling the time series}\label{sec3.4}
\begin{figure}
            \centering
            \includegraphics[width=7cm]{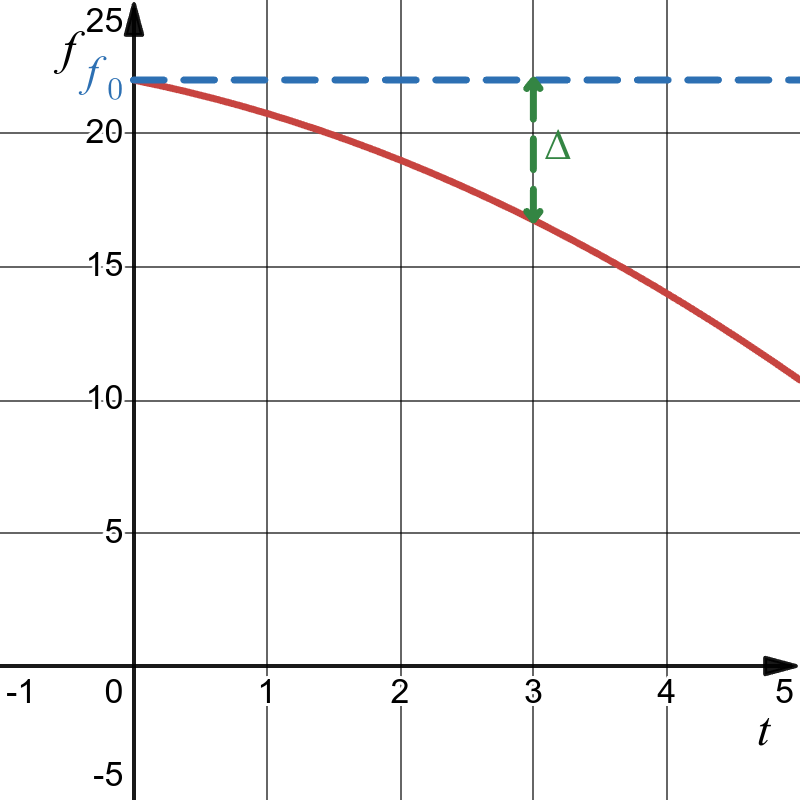}
            \caption{An example sketch of the signal frequency (inverse of the period) during the course of an observation. The blue dashed line is the constant frequency line at the initial frequency, $f_0$. $\Delta$ is the correct required to bring the signal to a constant frequency.}
            \label{fig:f_sketch}
        \end{figure}

        The deviation of the pulsar signal from strict periodicity is corrected for by transforming the time intervals of each time bin to account for the time-dependant frequency, $f$ (defined as the inverse of the pulse period, $p$; $f = 1 / p$.). We restrict the search to the frequency derivative, $f_1$ and double-derivative, $f_2$, assuming that the third derivative, $f_3$ is zero. If the observed time series has a pulsed signal of frequency given by 
        \begin{equation}
            f = f_0 + f_1 t + f_2 t^2 / 2,
        \end{equation}
        where $f_0, f_1$ and $f_2$ are the frequency and its derivatives at the beginning of the observation, respectively, then the correction to be applied to make the frequency constant, $f_0$ is given by (see \figref{fig:f_sketch} for a schematic diagram)
        \begin{equation}
            \Delta = f_0 - f = - f_1 t - f_2 t^2 / 2.
        \end{equation}
        The transformed sampling time interval $\tau$ is calculated according to the Doppler formula:
        \begin{equation}
            \tau = \tau_0\left(1 + \frac{f' t}{f_0} + \frac{f'' t^2}{2 f_0}\right),
        \end{equation}
        where $\tau_0$ is the original sampling time interval, and $f' = -f_1$ and $f'' = -f_2$ are the frequency derivatives to be applied as correction as explained below. The power in a particular time bin of the resampled time series is calculated by summing fractional powers from the corresponding bins in the time series to be resampled.
        
        \begin{figure}%
            \centering
            \subfloat[]{\includegraphics[width=7cm]{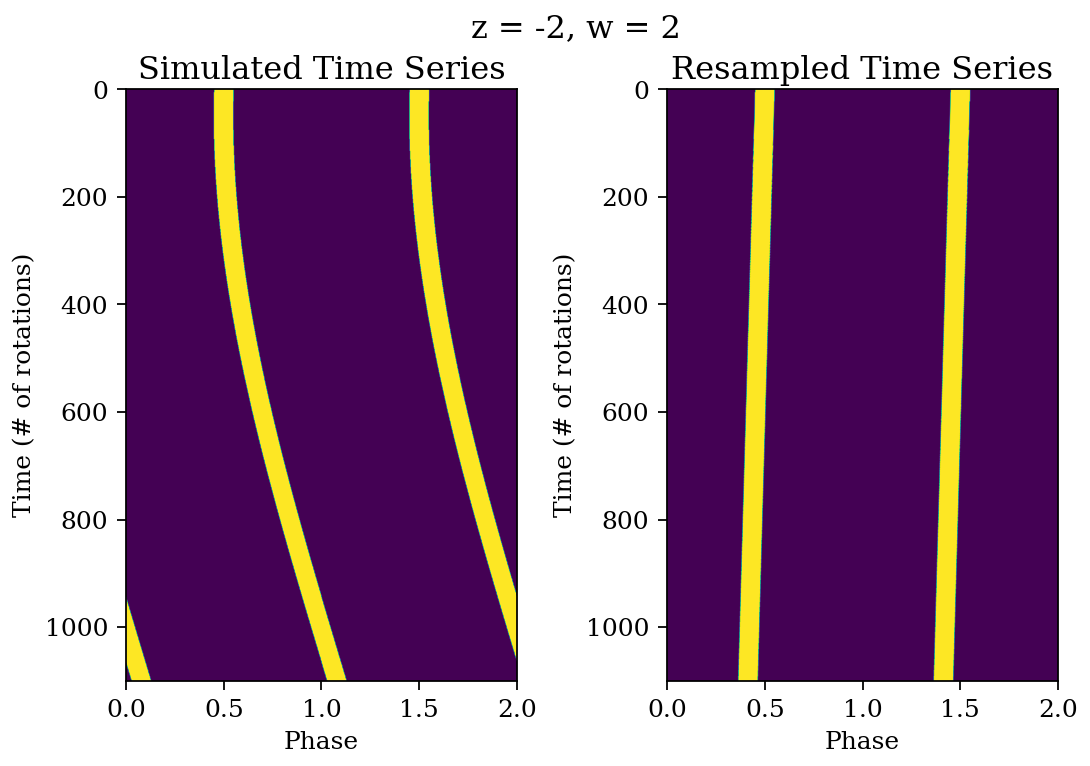}}
            \qquad
            \subfloat[]{\includegraphics[width=7cm]{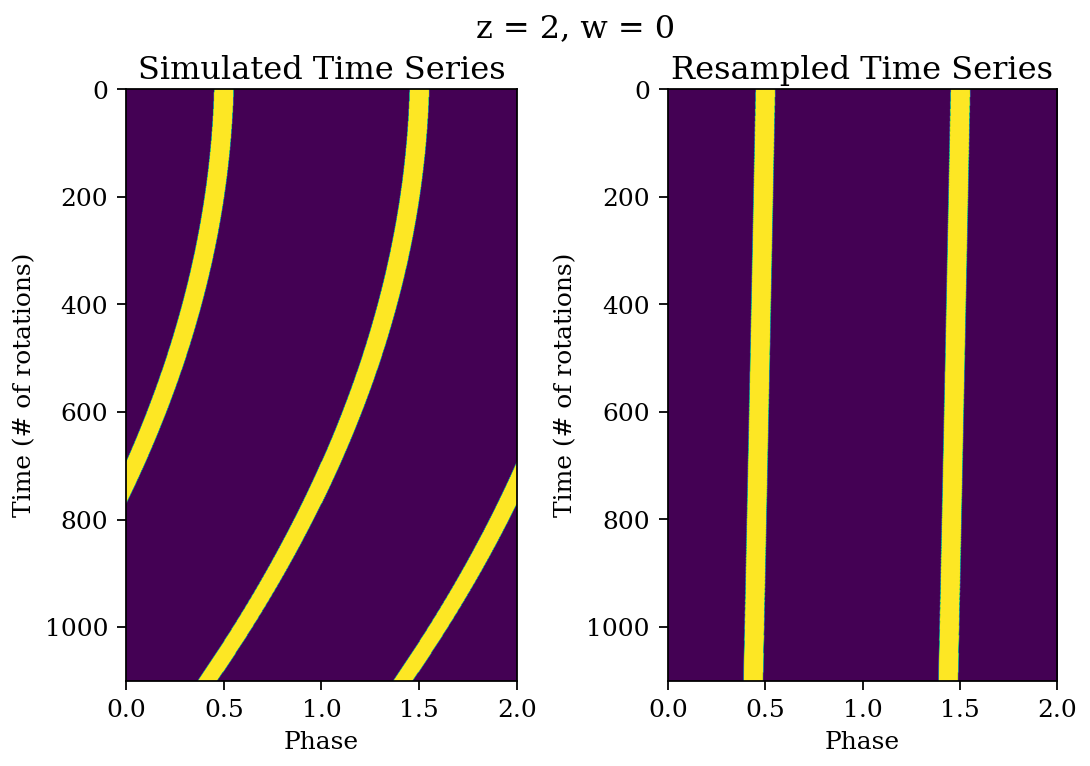}}\\%
            \subfloat[]{\includegraphics[width=7cm]{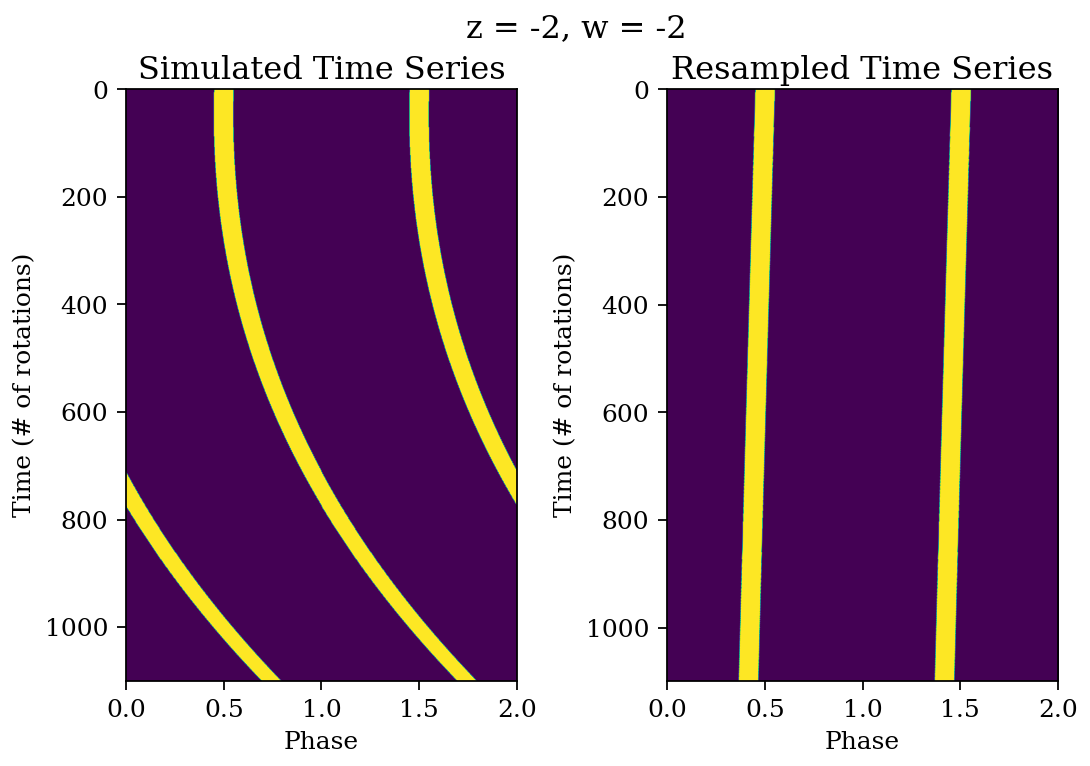}}%
            \qquad
            \subfloat[]{\includegraphics[width=7cm]{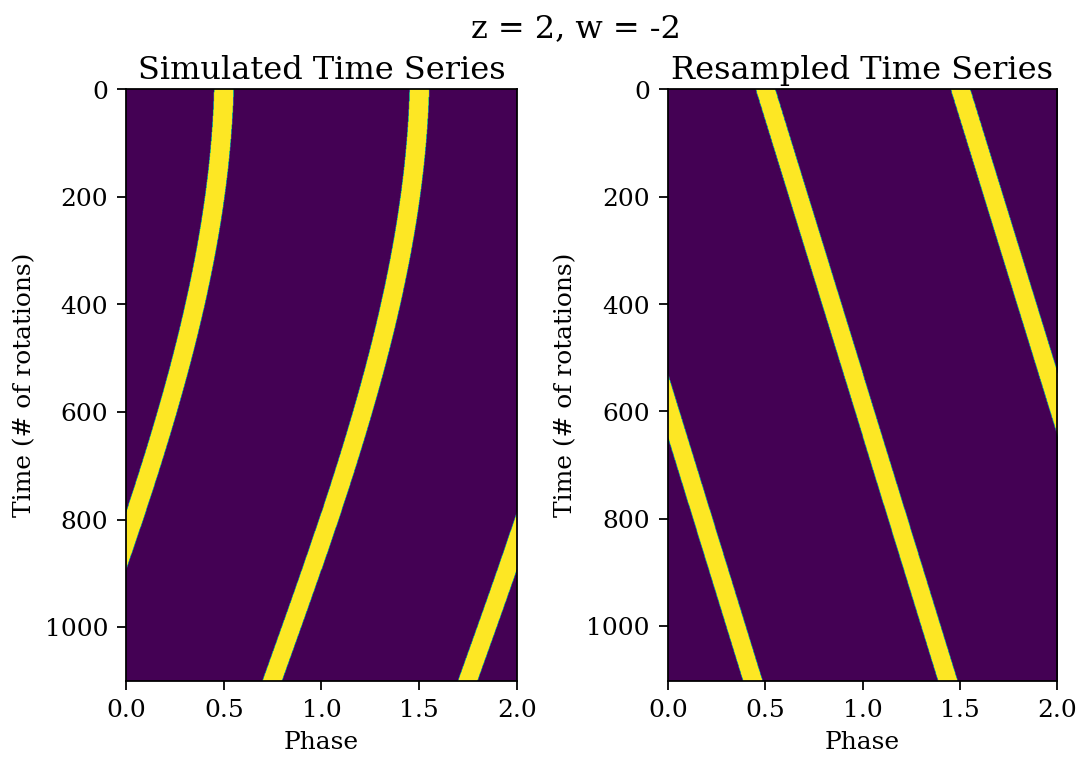}}%
            \caption{The left panel in all the sub-figures show the simulated boxcar pulse train with different frequency derivatives parametrised using z $= f_1 / T^2$ and w $= f_2 / T^3$. The right panel in all the sub-figures shows the corresponding resampled time series. The slope of the lines is due to fractional miss-match in the period due to finite sampling time intervals, and is between 1 and -1 sample per row as one would expect.}%
            \label{fig:resampling}%
        \end{figure}
        
        In the current version of the code, we assume that the frequency derivative does not change sign during the observation to avoid the complications of compressing a part of the time series and stretching the remaining part of it. If $f_1 < 0$, then $f$ decreases and $p$ increases, which implies the observed time series has to be compressed to constant period. In that case, we apply the above transformation as it is. In this case, the resampled time series will have a pulsed signal of a constant frequency $\tilde{f_0} = f_0$. If, however, $f_1 > 0$, then $f$ increases and $p$ decreases, which implies the observed time series has to be stretched to constant period. In this case, instead of stretching the time series, we reverse the time series so that it now has a signal with frequency
        \begin{equation}
            \tilde{f} = \tilde{f_0} + \tilde{f_1} t + \tilde{f_2} t^2 / 2,
        \end{equation}
        where 
        \begin{equation}
            \begin{split}
                \tilde{f_0} &= f_0 + f_1 T + f_2 T^2 / 2,\\
                \tilde{f_1} &= - f_1 - f_2 T\ \mathrm{and}\\
                \tilde{f_2} &= f_2
            \end{split}
        \end{equation}
        (since the third derivative is assumed to be zero) for an observation of duration $T$. We then apply the above transformation with
        \begin{equation}
            \begin{split}
                f' &= -\tilde{f_1} = f_1 + f_2 T\ \mathrm{and}\\
                f'' &= -\tilde{f_2} = -f_2.
            \end{split}
        \end{equation}
         After resampling the time series, we reverse the time series once again so that it then has a signal at constant $f = \tilde{f_0}$ instead of $f_0$. \figref{fig:resampling} demonstrates the resampling transformation on simulated train of boxcar pulses with different frequency derivatives.
         
    \section{Folding the time series}\label{sec3.5}
\begin{figure}
            \centering
            \includegraphics[width=10cm]{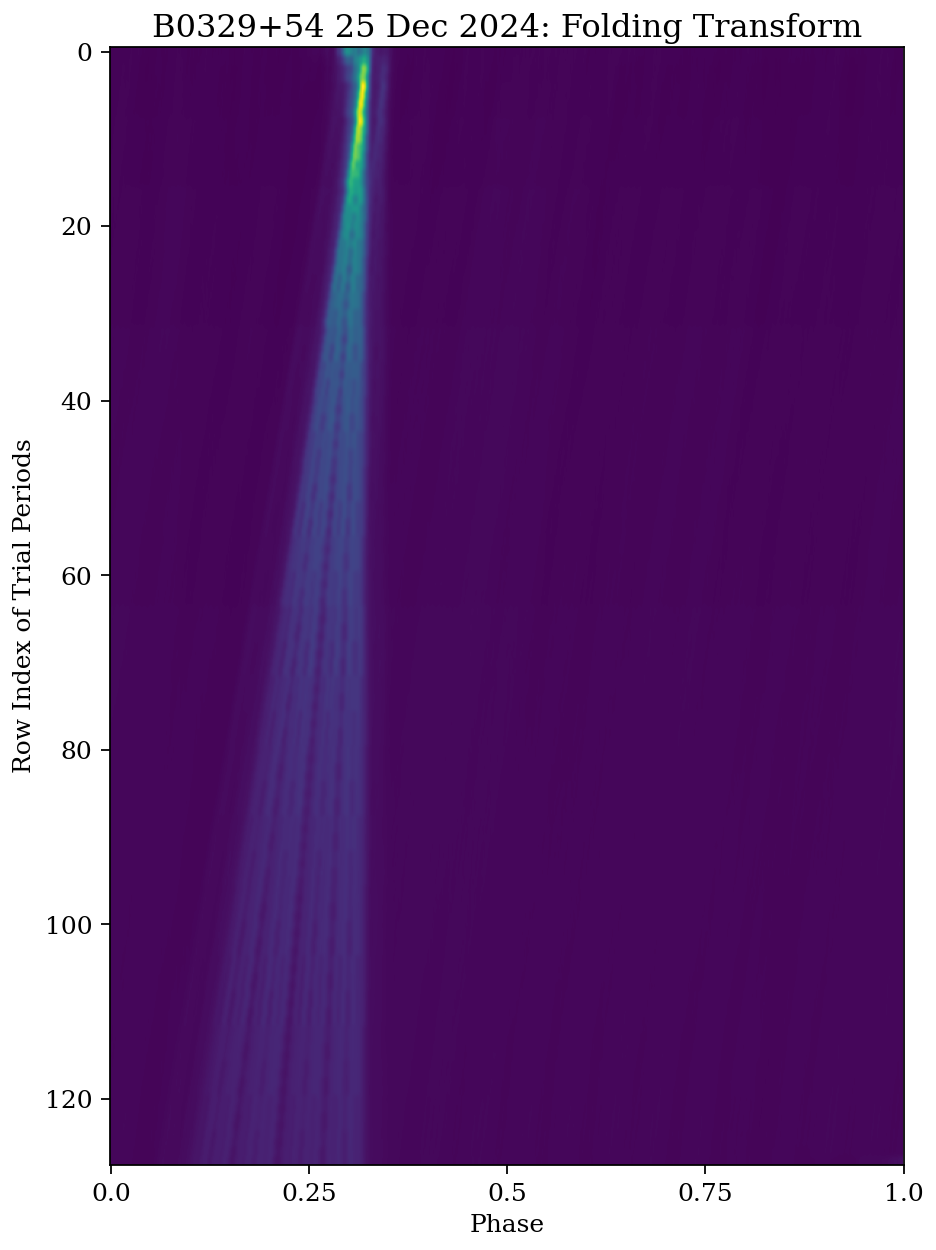}
            \caption{The folding transform of the same B0329+54 pulsar observation mentioned earlier. The base frequency was deliberately chosen to be only slightly less than its known value. The highest SNR (represented by the colouring scheme where yellow is the highest and violet is the lowest) thus appears near the top of the stack.}
            \label{fig:folding_transform}
        \end{figure}

        A time series of $N$ time samples containing a periodic pulsed signal of period $p$ time samples can be arranged as a 2-D stack of $m = \lfloor (N / p) \rfloor$ complete rows, where $\lfloor\ \rfloor$ denotes the floor function. The FFA folds such a stack to generate a stack of folded profiles at different trial periods in the range $p$ to $p^2 / (p - 1)$, i.e., the so-called folding transform (see \figref{fig:folding_transform}). The trial frequencies of the FFA are linearly spaced - the trial frequency, $f_s$ of the s-th row in the folding transform is given by
        \begin{equation}
            f_s = \left( \frac{1}{p} - \frac{s}{(m - 1) p^2} \right) \times \frac{1}{\tau_0}
        \end{equation}
        and the trial period is thus given by
        \begin{equation}
            p_s = \frac{p^2}{p -  s / (m - 1)} \times \tau_0.
        \end{equation}
        As is evident above, the trial frequency resolution improves with increasing $m$ and decreasing $\tau_0$. Thus, one has to pay a trade-off between better trial frequency resolution and computation time which increases with increasing $m$ and decreasing $\tau_0$. The folded profiles have pulses of different SNRs. A simple boxcar filter or more complicated pulse profile models can then be used to detect a statistically significant pulse in the folded profile. Dividing a long duration observation into smaller sub-integrations, the time evolution of the folded profile can be retained. Sub-banding of the dynamic spectrum, obtaining the sub-band-wise time series and doing the above mentioned analysis of all the sub-bands, the frequency evolution of the folded profile can be retained. A folded profile data cube of the same format as the PRESTO ``.pfd" format can be obtained which can then be passed on to the classifier.
        
    \chapter{Concluding Remarks}\label{ch4}
    Realising the need for high-throughput time-domain radio astronomy software for the next generation surveys at upcoming radio telescopes across the globe, SPOTLIGHT survey is as much an effort to fulfil this need as it is to unleash the transients and pulsar survey capabilities of a highly sensitive radio interferometer - GMRT. With the constant efforts of all the members of the collaboration, the survey is about to commence operations from the upcoming GMRT observation cycle 48 (from April 17, 2025 to September 29, 2025) in the commensal, targetted and open-sky modes.
    
    All the systems and software are undergoing rigorous testing for stability, functionality and performance. The real-time multi-beam transient search pipeline has already been thoroughly tested and is ready for operations, although without a real-time multi-beam RFI mitigation software. A data recording system along with an offline transient search pipeline is also ready and tested as a backup in case of any unexpected failure. An offline pulsar search pipeline performing acceleration search is also ready for operations with the folding step executing on CPUs until the GPU accelerated version is fully developed. The web-based control and monitoring interfaces, including the discovery database, are also ready for operations.
    
    There are several directions of future progress, especially along the lines of pulsar search. Boxcar filtering-based single pulse search on the folded profiles has to be integrated into the folding software. Multiple functions of the folding code needs to be accelerated on GPUs. The software has to undergo rigorous functionality and performance testing drawing comparisons with other software like PRESTO and RIPTIDE (\cite{morello_optimal_2020}). The output data format has to be adjusted to match the desired input format compatible with the classifier.
    
    The software developments are expected to happen parallel to exciting discoveries of FRBs and pulsars with up to arc-second localisation. A rich discovery database from SPOTLIGHT is anticipated to enable frontier research in FRB science using statistical techniques to study FRB-host galaxy environments and progenitor properties.
    
    \printbibliography
    \addcontentsline{toc}{chapter}{Bibliography}
    
\end{document}